\documentclass[fleqn,usenatbib]{mnras}
\pdfoutput=1
\usepackage[T1]{fontenc}

\DeclareRobustCommand{\VAN}[3]{#2}
\let\VANthebibliography\thebibliography
\def\thebibliography{\DeclareRobustCommand{\VAN}[3]{##3}\VANthebibliography}

\usepackage{graphicx}
\usepackage{amsmath}
\usepackage{amssymb}
\usepackage{booktabs}
\usepackage{multirow}
\usepackage{listings}
\usepackage[dvipsnames]{xcolor}
\usepackage[export]{adjustbox}
\usepackage{xspace}
\usepackage{makecell}
\usepackage{siunitx}
\usepackage{bm}
\usepackage{orcidlink}

\newcommand{\gmag}{\ensuremath{G}}
\newcommand{\gmagabs}{M_{\gmag}}

\newcommand{\gbp}{G_{\text{BP}}}

\newcommand{\grp}{G_{\text{RP}}}

\newcommand{\gcolor}{\gbp-\grp}
\newcommand{\gcolorabs}{\ensuremath{\left(\gcolor\right)_{0}}}

\newcommand{\deltamag}{\Delta\gmagabs}
\newcommand{\deltacol}{\Delta\gcolorabs}

\newcommand{\feh}{[\text{Fe}/\text{H}]}

\newcommand{\logg}{\log g}
\newcommand{\Teff}{T_\text{eff}}

\newcommand{\fbin}{\ensuremath{f_\text{bin}}}

\newcommand{\Rmax}{R_\text{max}}
\newcommand{\rmax}{r_\text{max}}
\newcommand{\zmax}{z_\text{max}}
\newcommand{\zmin}{z_\text{min}}

\newcommand{\gaia}{\textit{Gaia}\xspace}

\newcommand{\Msun}{\ensuremath{\text{M}_\odot}}

\newcommand{\logtyr}{\ensuremath{\log(t[\text{yr}])}}
\newcommand{\logtyri}[1]{\ensuremath{\log(t_{#1}[\text{yr}])}}

\DeclareMathOperator{\sech}{sech}
\DeclareMathOperator{\sechsq}{\sech^{2}}

\usepackage{newtxtext,newtxmath}

\title[Dissecting the \gaia HR diagram II]{Dissecting the \gaia HR diagram  II. The vertical structure of the star formation history across the Solar Cylinder}

\author[Mazzi et al.]{Alessandro Mazzi$^{1}$\thanks{E-mail:  alessandro.mazzi@hotmail.com} \orcidlink{0000-0002-7503-5078},
L\'eo Girardi$^{2}$ \orcidlink{0000-0002-6301-3269},
Michele Trabucchi$^{1}$ \orcidlink{0000-0002-1429-2388},
Julianne J. Dalcanton$^{3,4}$ \orcidlink{0000-0002-1264-2006},\newauthor
Rodrigo Luger$^{3}$,
Paola Marigo$^{1}$ \orcidlink{0000-0002-9137-0773},
Andrea Miglio$^{5,6,7}$\orcidlink{0000-0001-5998-8533},
Guglielmo Costa$^{8}$ \orcidlink{0000-0002-6213-6988},\newauthor
Yang Chen$^{9,1,10}$,
Giada Pastorelli$^{2}$ \orcidlink{0000-0002-9300-7409},
Morgan Fouesneau$^{11}$ \orcidlink{0000-0001-9256-5516},
Simone Zaggia$^{2}$ \orcidlink{0000-0001-6081-379X},\newauthor
Alessandro Bressan$^{12}$ \orcidlink{0000-0002-7922-8440},
Piero Dal Tio$^{1,2}$ \orcidlink{0000-0002-0834-5092}\\
$^{1}$Dipartimento di Fisica e Astronomia Galileo Galilei, Universit\`a di Padova, Vicolo dell'Osservatorio 3, I-35122 Padova, Italy \\
$^{2}$Osservatorio Astronomico di Padova -- INAF, Vicolo dell'Osservatorio 5, I-35122 Padova, Italy \\
$^3$Center for Computational Astrophysics, Flatiron Institute, 162 Fifth Ave, New York, NY 10010, USA\\
$^4$Department of Astronomy, Box 351580, University of Washington, Seattle, WA 98195 \\
$^{5}$ Dipartimento di Fisica e Astronomia, Università degli Studi di Bologna, Via Gobetti 93/2, I-40129 Bologna, Italy\\
$^{6}$ INAF - Osservatorio di Astrofisica e Scienza dello Spazio di Bologna, Via Gobetti 93/3, I-40129 Bologna, Italy\\
$^{7}$ School of Physics and Astronomy, University of Birmingham, Edgbaston, Birmingham, B15 2TT, UK\\
$^8$Univ Lyon, Univ Lyon1, Ens de Lyon, CNRS, Centre de Recherche Astrophysique de Lyon UMR5574, F-69230 Saint-Genis-Laval, France\\
$^{9}$Anhui University, Hefei 230601, China\\
$^{10}$National Astronomical Observatories, Chinese Academy of Sciences, Beijing 100101, China\\
$^{11}$, Max Planck Institute for Astronomy, Königstuhl 17, 69117, Heidelberg, Germany \\
$^{12}$SISSA, via Bonomea 365, I-34136 Trieste, Italy}

\date{Accepted XXX. Received YYY; in original form ZZZ}

\pubyear{2023}

\begin{document}
\label{firstpage}
\pagerange{\pageref{firstpage}--\pageref{lastpage}}
\maketitle

\begin{abstract}
Starting from the \gaia DR3 HR diagram, we derive the star formation history (SFH) as a function of distance from the Galactic Plane within a cylinder centred on the Sun with a 200~pc radius and spanning 1.3~kpc above and below the Galaxy's midplane.
We quantify both the concentration of the more recent star formation in the Galactic Plane, and the age-related increase in the scale height of the Galactic Disc stellar component, which is well-described by power-laws with indices ranging from $1/2$ to $2/3$.
The vertically-integrated star formation rate falls from $(1.147 \pm 0.039)\times10^{-8}\, \Msun \text{yr}^{-1} \text{pc}^{-2}$ at earlier times down to $(6.2 \pm 3.0) \times10^{-9}\, \Msun \text{yr}^{-1} \text{pc}^{-2}$ at present times, but we find a significant peak of star formation in the 2 to 3 Gyr age bin.
The total mass of stars formed per unit area over time is $118.7 \pm 6.2\, \Msun \text{pc}^{-2}$, which is nearly twice the present stellar mass derived from kinematics within 1~kpc from the Galactic Plane, implying a high degree of matter recycling in successive generations of stars.
The method is then modified by adopting an age-dependent correlation between the SFH across the different slices, which results in less noisy and more symmetrical results without significantly changing the previously mentioned quantities.
This appears to be a promising way to improve SFH recovery in external galaxies.
\end{abstract}

\begin{keywords}
Hertzsprung–Russell and colour–magnitude diagrams -- solar neighbourhood
\end{keywords}

\section{Introduction}
\label{sec:intro}

Thanks to the \gaia DR3 catalogue \citep{gaiaEDR3}, the Milky Way (MW) is now the only galaxy where the 3D distribution of stars can be precisely mapped across distances of a few kiloparsecs with resolution of a few tens of parsecs. Such a precise mapping is changing the perspective for stellar population studies: our specific position inside the MW is no longer ``a complication'' in the analysis of global properties of the MW, but might eventually become an advantage, allowing us to fully understand the geometry of its stellar populations and dust -- hence allowing for a better understanding of the populations that we are seeing, projected in 2D, in external disc galaxies.

Deriving the geometry of nearby stellar populations from \gaia -- and especially their vertical distribution, perpendicular to the Galactic Plane -- is indeed the goal of many recent papers, \citep[e.g.][]{bovy17,li21,yu21,everall22,widmark22,robin22}. These works make use of a variety of different techniques, applied to particular stellar tracers from \gaia, often taking advantage of their rich kinematic information, and of the modelling of the MW gravitational potential.

In addition, the precise photometry of the \gaia catalogue enables the construction of the colour-absolute magnitude diagram (CAMD), thus providing crucial information on the stellar ages.
Fitting this diagram with theoretical models of stellar populations allows us to derive the star formation history (SFH) in selected volumes, as done by, e.g., \citet{gallart19b}, \citet{ruiz20}, \citet{alzate20} and  \citet{daltio21}, using different methods and different sample sizes.  There are also more complete attempts to derive both the SFH and structural parameters of the Galactic Disc via population synthesis and kinematic modelling aimed to reproduce Gaia data, as is the case of \citet{mor19}, \citet{syso22}, and \citet{robin22}.

In this work, we attempt a different kind of analysis of \gaia data to derive the SFH of the extended Solar Neighbourhood. We consider a sample of stars in a cylindrical volume spanning approximately $1300$~pc above and below the Galactic Plane, with its axis normal to the Plane and passing through the Sun.
We slice it in layers along its axis and fit \textit{simultaneously} the CAMD of all slices, using a modified version of the SFH-recovery method routinely applied to nearby galaxies.
Our main goal is to map the vertical distribution of stellar populations of different ages, in a self-consistent way, to infer their relative contributions to the local disc density and their vertical distribution as a function of age.
At the same time, we aim to deriving the integrated properties of the MW disc close to the Sun, as would be seen by a distant observer. Last but not least, we have the long-term goal of improving the tools available for the analysis of SFHs in galaxies, and checking the assumptions that have been widely adopted in the study of external galaxies.

The structure of this paper is as follows. Sect~\ref{sec:data} describes the \gaia DR3 data used in this work. Sect.~\ref{sec:models} describes how the stellar populations are modelled considering the selection effects present in \gaia. Sect.~\ref{sec:methods} describes the numerical method used to derive the SFH in every region analysed. Sect.~\ref{sec:results} then discusses the results in terms of the age and spatial distribution of the SFH results. Sect.~\ref{sec:correlation} attempts to further improve the results via a new method that assumes a spatial correlation in the SFHs at different locations. Sect.~\ref{sec:conclusions} summarises the main conclusions.

\section{Data}
\label{sec:data}

\subsection{Building the sample}
\label{sec:gaiadata}

The initial sample of stars we download from the \gaia Archive is composed of all sources located in a sphere of radius 1.5~kpc centred on the Sun.
We use the query presented in the Appendix~\ref{app:query} of the online supplementary material, with $\rmax=1500$~pc.

The sample collected in this way is initially filtered to include only stars that fall inside a cylinder of radius $\Rmax=200$~pc and height $H=2.6$~kpc, spanning $H/2=1.3$~kpc above and below the Galactic Plane. The $\Rmax=200$~pc ensures small variations in the mean properties of stellar populations across its $400$~pc total range of Galactocentric radius: indeed, the change in mean metallicities due to the radial metallicity gradient in the Galactic disc should be smaller than 0.014~dex \citep[assuming the radial gradient of $-0.07$ dex/kpc;][]{rolleston00}, and the local stellar density should change by less than 15 per cent \citep[given a scale length for the thin disc of $2600$~pc;][]{juric08}. The maximum height of $H/2=1.3$~kpc is large enough to include at least one scale height of the oldest disc populations \citep[approximately 900~pc; ][]{juric08,pieres20}, and small enough to ensure high completeness and small parallax errors (see Sect.~\ref{sec:pmsin} below).

An important consideration is that the Sun is not located exactly at the Galactic midplane. Studies consistently find a $15-20$~pc offset between the midplane and the Sun. For example, \cite{karim2017}, using Galactic disc tracer objects, estimated that $z_{\odot}=17\pm5$~pc. \cite{siegert2019} determined the solar offset to be $z_{\odot}=15\pm 17$~pc using $\gamma$-rays. \cite{joshi2022} used open clusters younger than $700$~Myr and found that the solar offset amounts to $z_{\odot}=17.0\pm0.9$~pc. \cite{griv2021} determined a value of $z_{\odot}=20\pm2$~pc from Type II Cepheids, slightly larger than the other estimates.
In this paper, we adopt the weighted average of these results, and use $z_{\odot}=17.7$~pc.
We then compute the Cartesian coordinates $(x,y,z)$ of the stars in our initial sample as
\begin{equation}
    \begin{cases}
        x = r \cos(b) \cos(l) \\
        y = r \cos(b) \sin(l) \\
        z = r \sin(b) + z_\odot = r \sin(b) + 17.7 \, \mathrm{pc}
    \end{cases}
\end{equation}
where $(l,b)$ are the Galactic coordinates of each source and $r$ is the distance of the source from the Sun (see Sec.~\ref{sec:dist_ext}). We keep in our sample only the stars inside the Solar Cylinder, which we define using the following conditions:
\begin{equation}
    \label{eq:cylinder}
    \begin{cases}
        \sqrt{x^2 + y^2} \leq \Rmax \\
        \left|z\right| \leq 0.5 H
    \end{cases}
\end{equation}

\begin{figure}
    \centering
    \includegraphics[width=0.90\columnwidth]{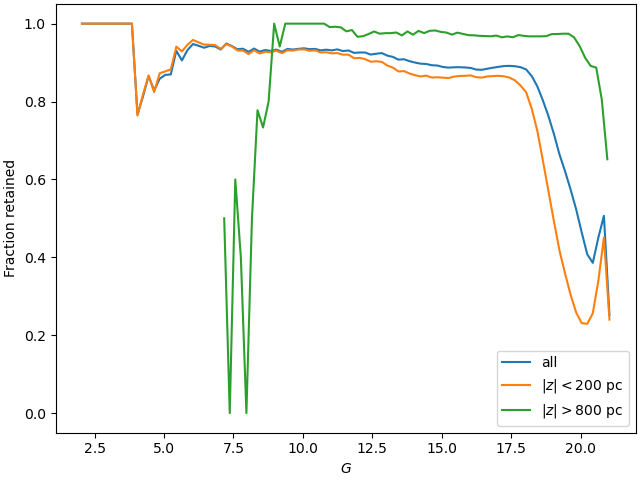}
    \caption{Fraction of stars retained in the sample after the cut on \texttt{phot\_bp\_rp\_excess\_factor} versus apparent $\gmag$ magnitude for different intervals of $z$: the full cylinder, $|z| < 1315.78~\mathrm{pc}$ (blue); $|z| < 200~\mathrm{pc}$ (orange); $|z| > 800~\mathrm{pc}$ (green). It should be noted that in the case $|z|>800$ only about 0.1 per cent of the stars have $\gmag<10$~mag.}
    \label{fig:stars_lost}
\end{figure}
After selecting stars in the Cylinder, we turn to the selection of stars based on the quality of their photometry.
\cite{riello2021} have shown that the \gaia photometric excess factor in the $\gbp$ and $\grp$ bands (\texttt{phot\_bp\_rp\_excess\_factor}, corrected using their eq.~6) is a measure of the consistency between the photometric measurements in the three \gaia bands, a key requirement for this study given the crucial role of the \gaia CAMD.
Therefore, we filter the sample according to section 9 of \cite{riello2021}. This filtering might eliminate high-amplitude variable stars and extended sources in addition to stellar sources with suspicious photometry.
Figure~\ref{fig:stars_lost} shows the fraction of stars retained, as a function of apparent \gmag-band magnitude in the whole cylinder, near the Galactic Plane and on the far sides of the Cylinder.
The bottom panel of Fig.~\ref{fig:camd} shows the Hess diagram, that is the 2D histogram of the CAMD, of sources lost over the whole cylinder.
To minimize the number of sources removed from the sample, we apply no other data quality cuts.

\subsection{Distance and extinction}
\label{sec:dist_ext}

We use the distances provided by the inverse of the \gaia DR3 parallax, since it represents a quite direct measurement with a clear definition of its error, which can be quite easily replicated in our models (see Sect.~\ref{sec:pmsin} below).

We computed parallax zero-points corrections following the prescriptions from \cite{lindegren21}\footnote{We use the Python package from \url{https://gitlab.com/icc-ub/public/gaiadr3_zeropoint}}. However, we found that the median value of the absolute change in parallax resulting from correcting our sample is less than 1 per cent, and that about 90 per cent of the stars in the sample have their parallaxes corrected by less than 2 per cent. Given the small impact of the correction and the way we include parallax errors in our models, we decided not to apply it.
This simplification is justified by the relatively small distance of all stars in our sample. The zero point correction could not be neglected were the sample to be extended over longer distances.

For the extinction, we adopt the 3D extinction map from \cite{vergely2022}. 
These authors devised a novel technique that allowed them to inter-calibrate different extinction catalogues and to obtain a 3D extinction density map via a hierarchical inversion technique with different volume and spatial resolution\footnote{We denote here ``spatial resolution'' the correlation length used by the authors to compute the map.}. In view of possible expansions of this work to larger distances, we chose the map that covers a volume of $6000~\mathrm{pc} \times 6000~\mathrm{pc} \times 800~\mathrm{pc}$ with a resolution of $25~\mathrm{pc}$. This map is available at the CDS\footnote{\url{https://cdsarc.cds.unistra.fr/ftp/J/A+A/664/A174/fits/}} and stores the extinction density at $550\mathrm{nm}$ ($A_{0}$). To determine the extinction of a star at a certain distance, we integrate the extinction density along the line of sight to the star and convert the value to \gaia magnitudes using the simple relation
\begin{equation}
    A_{\mathrm{X}} = f_{\mathrm{X}} A_{0}
\end{equation}
where $\mathrm{X}$ is one of $\gmag$, $\gbp$ and $\grp$, and
\begin{equation}
    \begin{split}
        f_{\gmag} &= 0.83627 \\
        f_{\gbp} &= 1.08337 \\
        f_{\grp} &= 0.63439 
    \end{split}
\end{equation}
are the values suitable for a yellow dwarf in the regime of low extinction, computed for the \gaia EDR3 photometric system as in \citet{girardi08}, using the \citet{odonnell} mean interstellar extinction curve with $R_V=3.1$. Then, we subtract the expected values of extinction from the magnitudes of the stars. In the whole sample, we find a maximum value of $A_{0}=0.89$~mag, and close to the Galactic Plane ($|z|<50$~pc) more than 75 per cent of the stars have $A_{0}<0.15$~mag. 

\begin{figure}
    \centering
    \includegraphics[width=0.90\columnwidth]{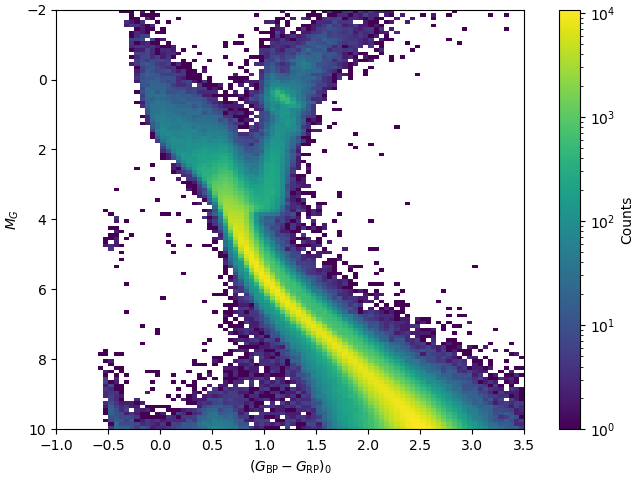}
    \includegraphics[width=0.90\columnwidth]{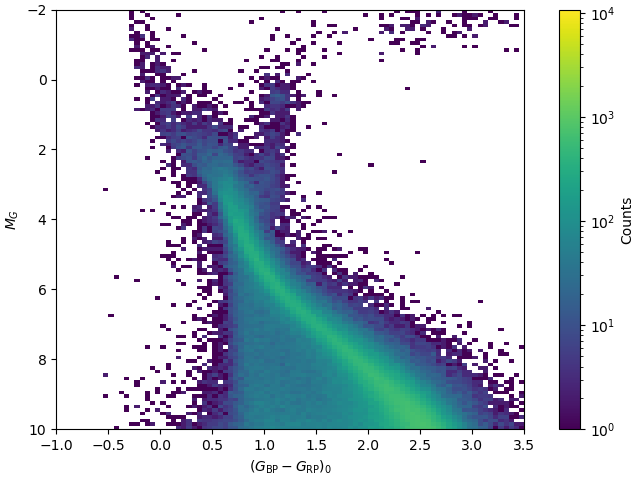}
    \caption{\textit{Upper panel}. Hess diagram of all the sources that were kept in the sample (5128792) based on the \texttt{phot\_bp\_rp\_excess\_factor} selection.
    \textit{Lower panel}. Same as above, but for all the sources that were removed from the sample (1594031) based on the \texttt{phot\_bp\_rp\_excess\_factor} selection. For both diagrams, the size of the bins is $0.1$~mag in magnitude and $0.05$~mag in colour. A clear feature can be seen in the upper panel at $3.5\lesssim\gmagabs\lesssim5$ and $\gcolorabs\approx-0.5$ which corresponds to the locus of the hot subdwarf stars, a product of binary evolution.}
    \label{fig:camd}
\end{figure}

The resulting CAMD is shown in the top panel of Fig.~\ref{fig:camd} in the form of a Hess diagram, that is, a two-dimensional histogram built from the colours and magnitudes of the stars in our sample.

\subsection{Slicing the data}
As we aim to determine the vertical structure of the SFH, we slice our cylindrical sample of stars into thin slices parallel to the Galactic Plane (hereafter referred to simply as ``slices''). More specifically, we split the sample in 28 slices with $\Delta z\approx 50$~pc for $\left|z\right|<158$~pc, and $\Delta z\approx 100$~pc for $\left|z\right|>158$~pc. The exact intervals in $z$ assigned to each slice are listed in Table~\ref{tab:slices}.

The reason why we chose this layout, with smaller heights near the Galactic Plane, is to improve the resolution at lower $z$, where the number of stars is substantially larger than at larger distances from the Plane of the Galaxy.

\begin{table}
    \centering
    \caption{Intervals of $z$ covered by each slice of the initial sample on the Northern side ($z>0$) of the Galactic Plane, such that $\zmin \leq z \leq \zmax$. The same scheme is repeated on the opposite side ($z<0$).
    }
    \label{tab:slices}
    \begin{tabular}{cc}
        \toprule
        $\zmin$ [pc] & $\zmax$ [pc]\\
        \midrule  
            0.00 &   52.63 \\
           52.63 &  105.26 \\
          105.26 &  157.89 \\
          157.89 &  263.16 \\
          263.16 &  368.42 \\
          368.42 &  473.68 \\
          473.68 &  578.95 \\
          578.95 &  684.21 \\
          684.21 &  789.47 \\
          789.47 &  894.74 \\
          894.74 & 1000.00 \\
         1000.00 & 1105.26 \\
         1105.26 & 1210.52 \\
         1210.52 & 1315.78 \\
          \bottomrule
    \end{tabular}
\end{table}

\section{Models}
\label{sec:models}

\subsection{Creating partial models for single stars}
\label{sec:pmsin}

Partial models (PM) are Hess diagrams of simple stellar populations covering limited ranges of age and metallicity.
In the case of single stars, they are built using the TRILEGAL code \citep{girardi05} with the same approach as in \citet[][see their fig.~4]{daltio21}: population models are computed for 7 values of mean metallicity around a reference age--metallicity relation (AMR), and for 16 age bins. The age bins are 0.2 dex wide in $\logtyr$, with the exception of the first age bin that spans 0.5 dex, overall providing a good balance between high enough statistics (even for the sparsely-populated slices far from the Galactic Plane) and age resolution.

The partial models are based on PARSEC v1.2S evolutionary tracks \citep{bressan12,chen14} and comprise the synthetic photometry in the \gaia DR3 filter transmission curves. They are also normalised to a constant star formation rate (SFR) of 1 \Msun\,yr$^{-1}$ over their age interval, assuming the \citet{kroupa02} initial mass function.
Examples of partial models for single stars built in this way are presented in Appendix~\ref{app:pms} of the online supplementary material.

As such, these initial partial models do not include the errors and incompleteness that characterise the real \gaia data. 
In the classical works of SFH analysis \citep[see e.g.][and references therein]{mazzi21} the completeness and photometric errors are derived via artificial star tests (ASTs): one injects huge numbers of fake stars in the original images and then recovers them using the same photometry pipeline used for the real data. Then, for every position in the colour-magnitude diagram, one derives the distribution of errors as the differences between the input and output colours and magnitudes, and the completeness as the number of recovered stars compared to the input fake stars. These quantities are a function of sky position. They are then applied to the partial models, which represent a complete sample of model stars.

It is obvious that the procedure has to be different for \gaia data, because its photometry and astrometry does not simply derive from CCD images, but from a complex observational procedure and data analysis pipeline \citep[see][and references therein]{gaiaEDR3,gaiaDR3}.
Fortunately, the errors and incompleteness of the \gaia DR3 catalogue have been characterised extensively in a series of papers. In the present analysis, we adopt the following procedure, which represents an extension and update of the one used by \citet{daltio21}:
\begin{enumerate}
    \item For every slice of the cylinder as defined in Table~\ref{tab:slices}, we generate random positions $(R,z)$ for 2 million fake stars with constant density within the limits of $R < \Rmax + \delta\Rmax$ and  $\zmin - \delta z < z < \zmax + \delta z$, where $\delta\Rmax=0.2 \Rmax$ and $\delta z = 0.1 (\zmax-\zmin)$. We generate fake stars in a volume larger than each slice because parallax errors open the possibility that stars generated within the $R<\Rmax$ and $\zmin<z<\zmax$ limits are ``observed'' slightly outside it; conversely, stars slightly outside the slice can be inserted into the sampled region. 
    We compute the parallax of every fake star from their distances from the Sun, and apply fake errors to its  parallax. These errors are derived from the parallax errors reported in the \gaia DR3 catalogue, for every slice, following the procedure outlined in Appendix \ref{app:errors} of the online supplementary material.
    
    \item Every fake star is assigned a random absolute magnitude and colour, within $-2<\gmagabs<12$ and $-1.2<\gcolorabs<4$ respectively. These limits are kept, intentionally, wider than the data we actually analyse in the end.
    The absolute magnitudes are then converted to apparent magnitudes using the fake parallaxes computed in (i) and are subsequently degraded applying the typical errors in $\gmag$, $\grp$ and $\gbp$ determined from the \gaia catalogue as outlined in Appendix \ref{app:errors} of the online supplementary material.
    
    \item The apparent magnitudes and sky coordinates of every fake star are provided as input to the GaiaUnlimited Python module for the completeness of the \gaia catalogue \citep{gaiaunlimited}, which computes the probability $P$ of observing each fake star in \gaia DR3. We use a randomly generated number $p$, between 0 and 1, to determine if the fake star is observed ($p<P$) or not ($p>P$).

    \item The fake star experiments presented above are used to derive the distributions of errors in absolute magnitude and colour and the incompleteness for every small bin in the CAMD. In particular:
    \begin{itemize}
        \item All stars generated inside the $R<\Rmax$ and $\zmin<z<\zmax$ volume and the $-2<\gmagabs<12$ and $-1.2<\gcolorabs<4$ CAMD limits are counted as ``input fake stars'', $n_\mathrm{input}$.
        \item All stars that are found, after their errors are simulated, inside the $R<\Rmax$ and $\zmin<z<\zmax$ volume and the $-2<\gmagabs<10$ and $-1<\gcolorabs<3.5$ CAMD limits are counted as ``observed stars'', $n_\mathrm{observed}$, and contribute to the estimation of the colour and magnitude errors.
        \item The completeness fraction is computed as $C = n_\mathrm{observed}/n_\mathrm{input}$.
    \end{itemize}
\end{enumerate}

We remark that the fake stars in the $\delta z$ and $\delta R$ regions allow us to estimate how much each partial model is affected by stars that are scattered inside/outside the sampled volume due to their parallax errors.
For a volume centred on the Sun, the probability of stars being scattered inwards is slightly higher than the probability of stars being scattered outwards\footnote{This is a subtle density bias closely related to the better-known Lutz-Kelker bias \citep{lutz} in the mean parallaxes.}. Therefore, this opens the possibility of finding completeness fractions $C$ exceeding the value of 1 in some cases. For the present choice of parameters, the maximum values of $C$ are of 1.007, and $C>1$ values are found in the partial models of slices closer to the Galactic Plane.

\begin{figure}
    \includegraphics[width=0.90\columnwidth]{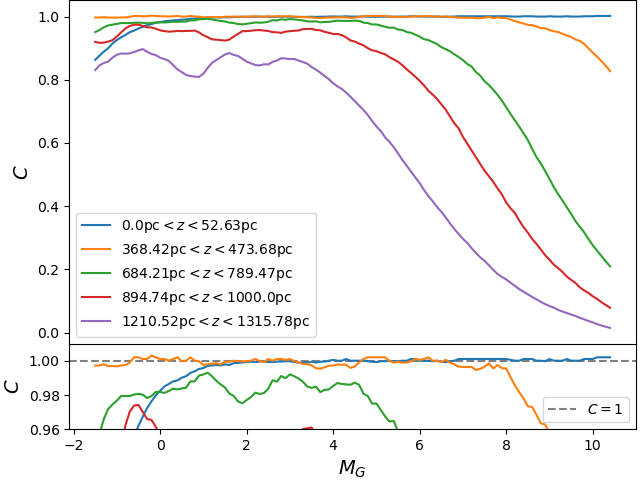}
    \caption{Results of the artificial star tests in five slices, illustrating the trend of completeness with the $\gmagabs$ magnitude. The lower panel zooms to the region with high completeness to highlight the occurrence of $C>1$ for the slices closest to the Galactic Plane.}
    \label{fig:completeness_profiles}
\end{figure}

Figure~\ref{fig:completeness_profiles} illustrates the results of such ASTs for a few cases: the slice closer to the Sun ($0.00~\text{pc}<z<52.63~\text{pc}$); three intermediate ones ($368.42~\text{pc}<z<473.68~\text{pc}$, $684.21~\text{pc}<z<789.47~\text{pc}$ and $894.74~\text{pc}<z<1000.00~\text{pc}$); and the most distant one ($1210.52~\text{pc}<z<1315.78~\text{pc}$).
In the first case, completeness is extremely high, except for a moderate loss of stars at $\gmagabs\lesssim 0$ because of their saturation\footnote{\gaia's detection efficiency drops for stars brighter than $\gmag\approx 3$ \citep{fabricius21}.
A star with absolute magnitude $\gmagabs=0$ has $\gmag=3$ at a distance $d\approx 40$~pc, which means stars at this absolute magnitude inside a sphere of radius $40$~pc are subject to saturation.
The volume of this region is approximately 2.1 per cent of the volume of the slice.
For $\gmagabs=-1$, $d\approx 63$~pc and the volume is 8.3 per cent of the slice's volume. These values look, overall, compatible with the $\approx5$ per cent loss in stars observed in Fig.~\ref{fig:completeness_profiles} for the slice closest to the Plane.}.
In the following cases, completeness progressively decreases until, in the furthest slice, no fewer than about 10\% of the stars are lost at all magnitudes, with this fraction increasing rapidly at $\gmagabs\gtrsim 4$. This limit is a consequence of the combination of incompleteness as determined by GaiaUnlimited and the increase in photometric errors for fainter sources.
In all the cases presented above, photometric errors are very small: for $\gmagabs<4$, we find a median magnitude error of $\sigma_{\gmagabs}=0.021$~mag and 99 per cent of the artificial stars have $\sigma_{\gmagabs}<0.102$~mag. Similarly, the median error on the colour is $\sigma_{\gcolorabs}=0.0034$, with 99 per cent of the artificial stars having $\sigma_{\gcolorabs}<0.0141$.

Finally, these distributions of errors and completeness are applied to the original, ``input partial models'', generating realistic ``observed partial models'' for every slice.

\subsection{Creating the partial models for binaries}
\label{sec:pmbin}
For binaries, the above-mentioned procedure gains a new complication: binary components too close on the sky and with a too large contrast in magnitude are identified as a single star in the \gaia DR3 catalogue. This effect depends on the distance from the Sun, with distant samples having a larger fraction of unresolved binaries than nearby samples.

To take this factor into account, we adopt the same procedure as in \citet{daltio21}: the BinaPSE module of the TRILEGAL code is used to generate coeval samples of binaries starting from the initial distribution of orbital parameters derived by \citet{moe17},
and using the \citet{kroupa02} IMF for the primary masses.
These are our ``binary partial models'': they have the same initial total mass as the single star partial models of the same age and metallicity and are normalised to a constant SFR of 1 \Msun\,yr$^{-1}$ as well.
Then, BinaPSE follows the evolution of the binaries, both in their physical properties and in their orbital parameters.
At the selected partial model age, we generate the magnitude contrast between the two components, as well as the expected separation of the binary members for a random epoch, random orientation of the orbit (i.e., for an isotropic distribution), and random position  of the centre of mass inside the slice.
If the combination of angular separation and magnitude contrast between the components turns out to be larger than the limits defined by eq.~2 of \citet{smart21} (which replaces the \citealt{ziegler18} and \citealt{brandeker19} relations based on \gaia DR2), they are treated as two separated stars, each one with its own photometry. Otherwise, they are treated as a single object, with the photometry being computed by adding the fluxes of the two components. (or by the left-over star in the case of coalesced binaries). In both cases, the photometric errors and completeness already derived from single stars are applied, to convert the input binary partial models into the final observed ones.
Examples of binary partial models are provided in Appendix~\ref{app:pms} of the online supplementary material.

It is worth pointing out the importance of accounting for both resolved and unresolved binary stars, especially when using \gaia data, even though they are often neglected in literature studies dealing with SFH recovery.
While the detailed implications will be addressed in a forthcoming paper on the binary fraction of the Solar Neighbourhood, we can work out a simple instructive example comparing a binary partial model in a slice close to the Plane and one at 1~kpc.
Indeed, the number of objects in the latter one is about 6 per cent smaller than the former one, even though they refer to the same population of binaries.
If not properly accounted for, the effect of resolvability is prone to propagating into the values derived for the absolute SFR and the binary fraction.
Even more subtle effects may derive from the fact that the fraction of some binaries is a function of orbital separation. According to \citet{moe17}, this happens for the ``twins'' -- that is, the binaries in which the two components have a similar initial mass -- which are favoured at small periods. Since unresolved twins have a very distinct location in the CAMD, the incorrect simulation of their resolvability at different distances may also affect the residuals of the model fits across the CAMD.

\subsection{Creating a partial model for the Galactic halo}
\label{sec:halo}
It is well known that the Solar vicinity is immersed in a low-density halo of old metal-poor stars, that can be followed spatially up to very large distances \citep[e.g.][]{juric08}, and which is responsible for a background of deep star counts over large areas of the sky \citep[e.g.][]{pieres20}.
This background of stars is expected to vary modestly and smoothly across our Solar Cylinder, and we can consider it as a fixed element in the fit of the Hess diagram.
In this work, we adopt the TRILEGAL default halo model \citep{girardi12} to create the halo population for every slice, and then apply the corresponding ASTs as described above, creating additional partial models called PM$_0$.

\subsection{Computing the total model and its likelihood}
From the partial models described above, we can build a so called ``total model'' for a slice as
\begin{equation}
    \text{M} = \text{M}(\text{SFH}, \fbin, \deltamag, \deltacol) + \text{PM}_{0}
\end{equation}
where SFH is the star formation history of the slice, $\fbin$ is the binary fraction, which we define as the mass fraction of a stellar population initially born in the form of binaries, and $\deltamag$ and $\deltacol$ are global shifts of the Hess diagram intended to take into account possible issues in the \gaia photometry or in the definition of the filters of the model.
The SFH consists of two parts, the SFR as a function of age and the AMR, which describes how metallicity varies with age. We indicate by $a_{i}$ the value of the SFR in the $i$-th age bin, and by $\feh_{i}$ the corresponding value of metallicity.
Finally, $\text{PM}_{0}$ takes into account any additional contributions to the total Hess diagram that has to be kept fixed during the optimisation.
In our case, as discussed in Sec.~\ref{sec:halo}, we choose to keep fixed the partial model of the Halo.
However, in our testing, we find that including or excluding the Halo model does not change the result significantly.

In the computation of the total model there are a few intermediate steps. First, we need to derive the models with the requested metallicity. We do so by interpolating between the sets of both single and binary partial models we computed for different metallicities. More specifically, we get the partial models at age $i$ for the desired metallicity $\feh_{i}$ by interpolation:
\begin{equation}
    \label{eq:pm_metallicity}
    \mathrm{PM}_{i}(\feh_{i}) = (1-f_{i}) \mathrm{PM}_{i}^{-} + f_{i} \mathrm{PM}_{i}^{+}
\end{equation}
where $\mathrm{PM}_{i}^{-}$ and $\mathrm{PM}_{i}^{+}$ are the partial models at metallicity $\feh_{i}^{-}$ and $\feh_{i}^{+}$ respectively and at the age $i$ that bracket the target metallicity, and $f_{i}=(\feh_{i}-\feh_{i}^{-})/(\feh_{i}^{+}-\feh_{i}^{-})$. This allows us to compute a limited number of sets of partial models, thus reducing the time needed initially, but still be able to simulate the whole metallicity range.
The second step consists in combining single and binary models at each age.
Our definition of binary fraction $\fbin$ implies that at all ages and metallicities a SFR of 1 \Msun\,yr$^{-1}$ is represented by the sum of $(1-\fbin)$ times the partial model for single stars, plus $\fbin$ times the partial model for binaries.
Lastly, we combine the partial models at each age using the SFRs, $a_{i}$, as weights to obtain the total model M. Putting these last two steps together, the total model is computed as
\begin{equation}
    \text{M} = \text{PM}_{\text{0}} + \sum_{i=1}^{16} a_{i} \left[ (1-\fbin)\text{PM}_{\text{sin},i} + \fbin \text{PM}_{\text{bin},i} \right],
\end{equation}
where $\text{PM}_{\mathrm{sin},i}$ and $\text{PM}_{\mathrm{bin},i}$ are the partial models for single and binary stars, respectively.

The main objective of our work is determining the set of parameters that minimises the residuals between the model and the data for each slice of the cylinder.
To evaluate how well a model produced by a given set of parameters compares to the data, we adopt a method similar to the ones presented in \cite{daltio21} and \cite{mazzi21}.
We adopt 16 age bins, from $\logtyr=6.6$ to $\logtyr=10.1$, and compute models at those ages for a reference AMR (see figure~4 in \citealt{daltio21} and Fig.~\ref{fig:RAMR}) and six additional sets at $\lbrace-3,-2,-1,1,2,3\rbrace \Delta\feh$, where $\Delta\feh=0.12$~dex.
Every set has an intrinsic spread in metallicity defined by a Gaussian of $\sigma=0.1$~dex.

At a first glance, the parameters to fit for each slice are 16 $a_{i}$, 16 $\feh_{i}$, the binary fraction, and the two additional parameters describing the rigid shifts of the CAMD along the magnitude and colour axes, $\deltamag$ and $\deltacol$.
With 28 slices of data, the total number of parameters that have to be explored is 980.

To reduce the size of the problem, we can make a few assumptions:
\begin{itemize}
    \item In our Solar Cylinder sample we do not expect the binary fraction $\fbin$ to change with the height from the Galactic Plane, therefore we can fit a single value for all slices.
    \item Similarly, we do not expect $\deltamag$ and $\deltacol$ to change with $z$ and we can fit only one value of each parameter for the whole sample.
    \item For the AMR, the observed vertical metallicity gradient of the disc -- i.e. the trend of finding lower [Fe/H] values at high $|z|$ \citep[see, e.g., figure~4 in][and Appendix~\ref{app:AMR} of the online supplementary material]{hayden15} -- might be the simple result of having young, metal rich populations being confined at low $|z|$, while old metal poor stars dominate the star counts at high $|z|$. As we are exploring the age structure of the disc, we tentatively assume that its populations obey a single age--metallicity relation, and hence that any observed vertical metallicity gradient is the simple result of adding pieces of stellar populations which have different spatial distributions. This translates into a single choice for the AMR of all slices. This assumption is consistent with the observation of a narrow AMR in the Solar Neighbourhood \citep{feuillet18}. 
    \item Lastly, we adopt the same reference AMR as in \cite{daltio21}, linear with $\logtyr$, and implement the same simplification of \cite{daltio21} and \cite{mazzi21}. Instead of fitting the values of the AMR at each age, we choose to fit only the shifts in metallicity in the first and last age bin, $\Delta\feh_{1}$ and $\Delta\feh_{16}$, and compute the ones for the remaining age bins through linear interpolation. Further details about the reference AMR and its suitability are provided in Appendix~\ref{app:AMR} of the online supplementary material.
\end{itemize}
These choices result in the halving of the number of fitted parameters. In turn, having fewer parameters speeds up the convergence to the solution and should also reduce the chances of the solution falling into a local likelihood maxima\footnote{Although, as described in the following Section~\ref{sec:methods}, the algorithm we use for the fit should not be affected significantly by local maxima}.

To compare a model for a slice $n$ produced with a specific set of parameters $\theta_{n}$ to the data, we need to choose a likelihood function. As we are dealing with counts, the most immediate choice is a Poisson likelihood which, in its logarithmic form, is the following:
\begin{multline}
    \ln\mathcal{L}(\theta_{n}) = \sum_{j=1}^{N_{\mathrm{Hess}}} \Bigl( -\mathrm{M}_{n,j}(\theta_{n}) + \mathrm{O}_{n,j} \ln\bigl(\mathrm{M}_{n,j}(\theta_{n})\bigr) + \\
    - \ln(\mathrm{O}_{n,j}!) \Bigr)
    \label{eq:likelihood}
\end{multline}
where M and O are the Hess diagrams of the model computed with the parameters $\theta$ and of the observations, respectively, while $i=1,...,N_{\mathrm{Hess}}$ stands for the index of the cells of each Hess diagram.
The total likelihood is then obtained as the sum of the likelihoods of the single slices:
\begin{equation}
    \ln\mathcal{L}(\theta) = \sum_{n=1}^{N_{\mathrm{slice}}} \ln\mathcal{L}(\theta_{n})
\end{equation}

\subsection{Selection of the CAMD area}
\label{sec:camd_selection}
So far, we have defined and characterized the \gaia sample over a very ample magnitude interval, namely $-2<\gmagabs<10$ (see Fig.~\ref{fig:camd}). It is convenient however to limit the final CAMD analysis to a shorter range, so that:
\begin{itemize}
    \item We limit the uncertainties due to errors in the \gaia catalogue. Indeed, as seen in the bottom panel of Fig.~\ref{fig:camd}, a significant fraction of stars in the lower part of the CAMD occupy unrealistic positions, which might be attributed to errors in the astrometry.
    \item We maximise the sensitivity of our analysis to the stellar population age. Combined with the effect of completeness, this implies favouring the range $\gmagabs<4$, which comprises all main sequence turn-offs up to very old ages with a completeness of at least 90 per cent at all magnitudes included in the analysis.
    \item We limit the systematic errors in the stellar models that likely appear when we try to cover too wide a range in effective temperature and surface gravity. In fact, as shown in \citet{daltio21}, the simultaneous fit of the entire CAMD, including both unevolved low-mass stars and evolved massive stars, is quite problematic.
    \item We exclude the AGB stars, which usually appear as long-period variables excluded by our selection criterium in Sect.~\ref{sec:gaiadata}, and whose stellar models depend on a series of parameters which are actually calibrated using the SFHs of nearby galaxies of smaller metallicity than the MW disc \citep{pastorelli19,pastorelli20}, by taking $\gmagabs>-1$. 
\end{itemize}
Taking into account these considerations, in the following we limit the analysis to the $-1<\gmagabs<4$ interval. 

Another important aspect is the resolution of the Hess diagrams derived from the CAMD. A high resolution is needed to be able to identify the small scale features, but this has to be balanced by the requirement of a large enough number of sources in each bin. We find that a Hess diagram with a $0.1$~mag resolution in magnitude and $0.05$~mag resolution in colour achieves this balance, and we use these resolutions throughout this work.

\section{Methods}
\label{sec:methods}
\subsection{The fitting method}
We develop and employ a fitting method consisting of two separate phases.
In the first phase, our goal is to move as close and as reliably as possible to the solution.
An algorithm suited for this class of problems is gradient descent, which in its simplest form consists in moving from a starting guess towards the solution according to the direction dictated by the gradient of the objective function, therefore progressively lowering the residuals between the model and the data.
The distance by which the solution moves at each iteration is given by the product between the gradient for each fitted parameter and a hyperparameter called the \textit{learning rate}.
While in principle simple to implement, a method based on pure gradient descent can be slow if the number of parameters or the number of points where the gradient is computed are large.
The stochastic gradient descent algorithm improves the performances by computing the gradient at each iteration only for a randomly chosen sub-sample of the dataset. The method we chose for our optimisation is called \textit{Adam} \citep{adam} and it is an expansion of the stochastic gradient descent.
The fundamental advantage it has over the previous methods is that the learning rate for each fitted parameter is adjusted while iterating based on the first and second momenta of the gradient of each parameter, allowing a better automatic tuning of the learning rate hyperparameter and effectively a faster convergence to the solution.

After the optimisation of the parameters of the model with \textit{Adam}, we are generally close enough to the solution to be able to switch to the second phase of our method, consisting in a Markov chain Monte Carlo (MCMC) algorithm. This phase is aimed at further refining the solution and providing the confidence intervals of each parameter of the fit. Of all the algorithms available, we chose a variation of the Hamiltonian Monte Carlo (HMC) algorithm, namely the No-U-Turn Sampler  \citep[NUTS;][]{nuts}. HMC is useful as it attenuates the random walk often encountered in standard MCMC methods by performing sub-iterations according to the variations of the gradient of the objective function, however its performance has a strong dependence on its two hyperparameters: the step size and the number of steps to walk. For this reason, HMC often needs some hand-tuning to perform well. On the other hand, NUTS is capable of automatically adapting these hyperparameters, thus avoiding the need for accurate hand-tuning.

Our method is coded in Python, and we use the \texttt{numpyro} package\footnote{\url{https://github.com/pyro-ppl/numpyro}} \citep{phan2019numpyro,bingham2019pyro}, a ``lightweight probabilistic programming library'', that takes advantage of the \texttt{jax} library\footnote{\url{https://github.com/google/jax}} \citep{jax2018github} for automatic differentiation and just-in-time compilation.

We apply our method to the sample of data described in Sect.~\ref{sec:data}, consisting of a total of 28 slices spanning the $-1315.78~\text{pc} \leq z \leq 1315.78~\text{pc}$ range. In a typical run, we use 1000 iterations of the \textit{Adam} optimiser with a learning rate of $0.1$, and we find that increasing the number of iterations does not reduce the loss in most cases. In the subsequent MCMC phase, we make the chain walk 1000 steps, and we discard the initial 200 steps as warm-up.
The final result is then taken from the distribution of parameters determined from the MCMC run.
The errors on the best-fit parameters are derived from the steps of the MCMC. In particular, we take the chain after the warm-up phase and compute the $68^{\mathrm{th}}$ and $94^{\mathrm{th}}$ percentiles of the parameters as our confidence regions.

\subsection{Method validation}
\begin{figure*}
    \centering
    $0.00~\mathrm{pc} < z < 52.63~\mathrm{pc} $ \\
    \includegraphics[width=0.99\textwidth, trim={0 0 0 72px}, clip]{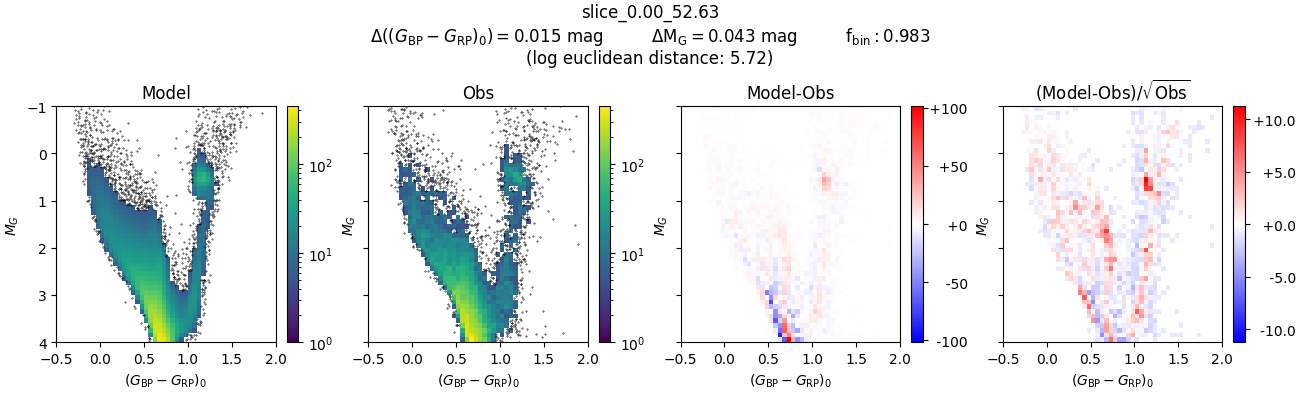}\\
    $368.42~\mathrm{pc} < z < 473.68~\mathrm{pc} $ \\
    \includegraphics[width=0.99\textwidth, trim={0 0 0 72px}, clip]{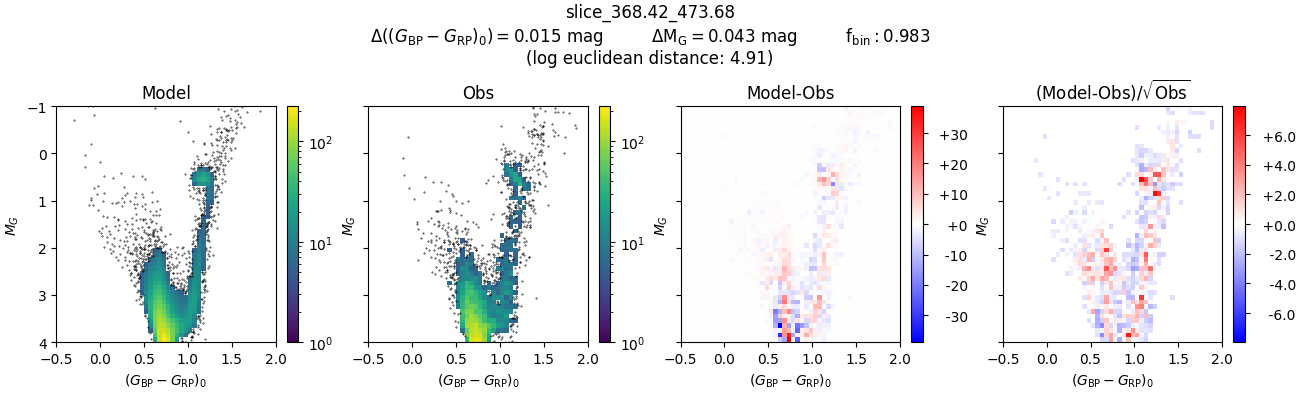}\\
    $684.21~\mathrm{pc} < z < 789.47~\mathrm{pc} $ \\
    \includegraphics[width=0.99\textwidth, trim={0 0 0 72px}, clip]{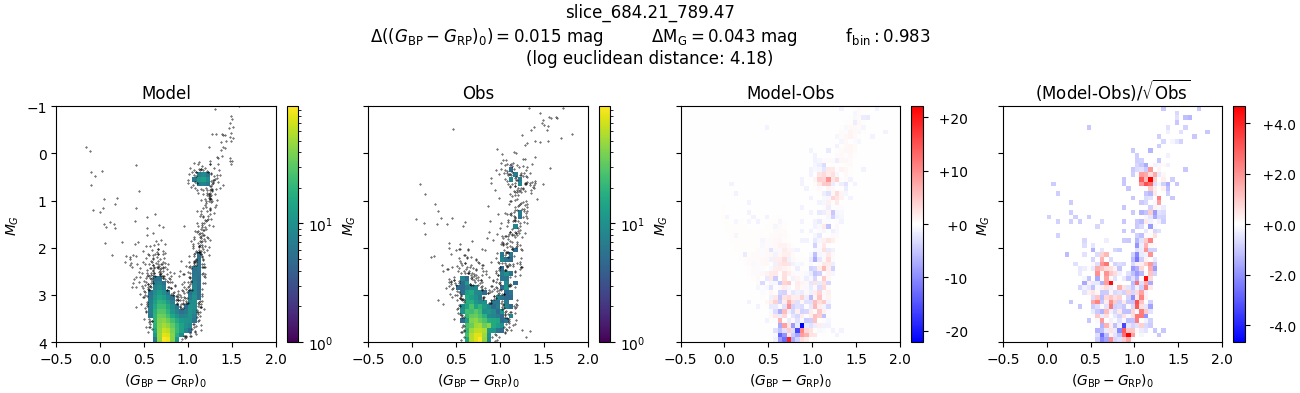}\\
    $894.74~\mathrm{pc} < z < 1000.00~\mathrm{pc} $ \\
    \includegraphics[width=0.99\textwidth, trim={0 0 0 72px}, clip]{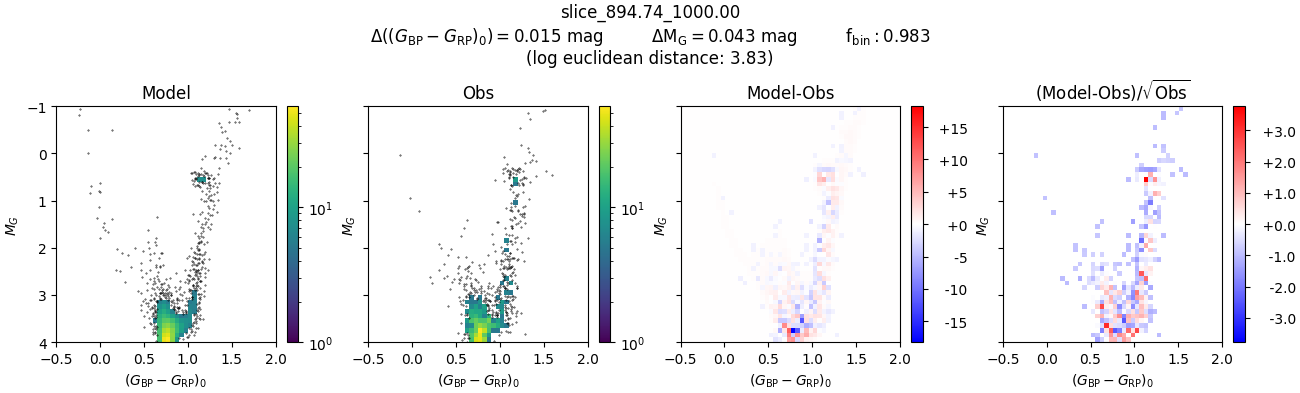}
    $1210.52~\mathrm{pc} < z < 1315.78~\mathrm{pc} $ \\
    \includegraphics[width=0.99\textwidth, trim={0 0 0 72px}, clip]{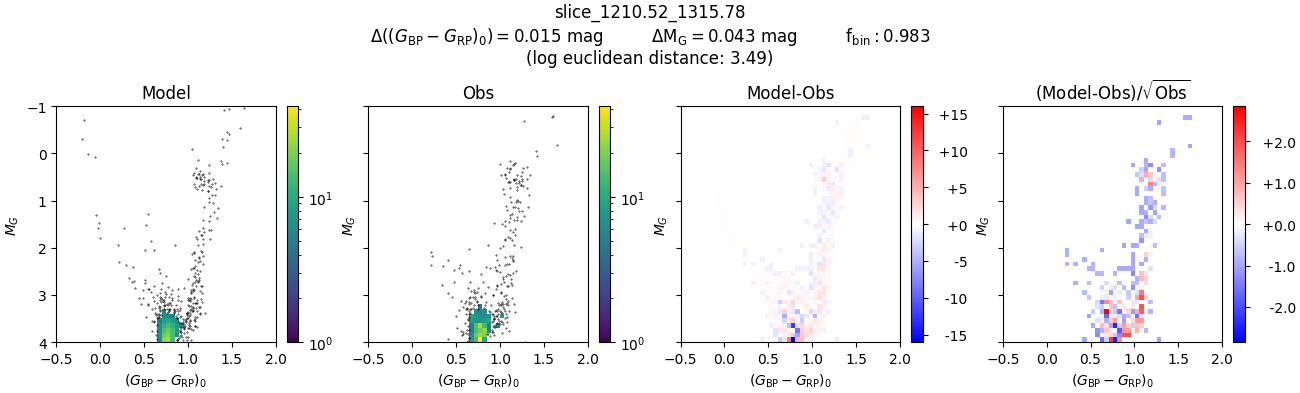}
    \caption{Comparison between the best-fit models and the observations in four slices of the sample, with distance from the Galactic Plane increasing from the top panel to the bottom panel. For every row, from left to right, the Hess diagrams correspond to: the model M produced with the best fitting parameters, the observations O, the residuals M-O and the significance (M-O)/$\sqrt{\text{O}}$}.
    \label{fig:results_comparison}
\end{figure*}
Figure~\ref{fig:results_comparison} shows examples of the fits for a few slices, comparing the best model with the observations and presenting their typical differences with their significance in the Hess diagram.
In the uppermost panel ($0.00\mathrm{pc}<z<52.63$~pc) of the first column, we note a small region with positive residuals around the Red Clump, as well as a zone with negative residuals upwards from the turn-off; in the bottom panel ($1210.52\mathrm{pc}<z<1315.78$~pc), the residuals are mostly concentrated around the turn-off. 
These kinds of residuals are similar to those found by \citet{daltio21} in the previous analysis of CAMDs from \gaia, and can be more easily detected in Hess diagrams produced with a high-resolution in colour and magnitude.
High residuals are also common in the fitting of the red clump region, even for nearby galaxies such as the Magellanic Clouds \citep[see e.g.][]{rubele18,mazzi21}. They are probably the consequence of small inadequacies in modelling the He-burning phase of low-mass stars, which are still to be fully understood \citep{girardi16}, not to say corrected.

The fitting method we use also produces a number of young stars than is larger that expected, as highlighted by an upper Main Sequence more extended towards bright magnitudes in the best fit model than in the observations. While possible sources for this issue will be discussed in the following section, we note here that the uncertainty on this young component is considerable and its contribution to the total mass is minimal.

Before discussing the SFH, we comment briefly on the other fit parameters. First, we find only small offsets in the observed versus model photometric parameters, namely $\deltamag = 0.04321 \pm +0.0009$~mag and $\deltacol = 0.01542 \pm -0.0002$~mag. These are reasonable given the known uncertainties in \gaia magnitudes and the difficulties in reproducing a new photometric system with synthetic photometry\footnote{These difficulties are made evident in the ``standardisation'' process adopted from \citet[][see their figure 4]{montegriffo}, who modify the transmission curves of well-established photometric systems to make the synthetic photometry of \gaia BP and RP spectra to reproduce the magnitudes of standard stars.}. Second, we find that $\Delta\feh_{1} = -0.0764 \pm 0.0008$~dex and $\Delta\feh_{16} = -0.083 \pm 0.001$~dex, therefore the AMR is shifted to slightly smaller metallicities than in the reference AMR.
Comparing our best fit AMR with the results of APOGEE and GALAH, shown in Fig.~\ref{fig:RAMR}, we note that they are generally not in contrast. However, a detailed verification would require the addition of spectroscopic data, with its own selection function to be taken into account, in our SFH-recovery method. This kind of analysis will be pursued in a future work in this series.

Finally, we find a high initial binary fraction, more precisely $\fbin = 0.984 \pm 0.003$. For the moment, suffice it to recall that a similarly high fraction of binaries has also been suggested by the frequency of the ``astrometric signals'' expected from binaries as compared to those observed in the \gaia sample of nearby stars \citep{penoyre22}. On the other hand, the broadening of the lower main sequence in \gaia CAMD seems to suggest \fbin\ values closer to 0.4 \citep{daltio21}. Both estimates strongly depend on the assumed intrinsic distribution of binary periods; however, there is a good agreement between the distributions assumed by the above-mentioned authors (based on \citealt{raghavan10} and \citealt{moe17}, respectively). Therefore, the real fraction of binaries appears undefined at the moment, with different subsamples of \gaia data and methods providing different estimates. Fortunately, our main results for the SFH depend little on the final value of \fbin. These aspects are discussed in more detail in Appendix~\ref{app:binary_uncert} of the online supplementary material.

 \section{Results and discussion}
 \label{sec:results}
\begin{figure}
    \centering
    \includegraphics[width=0.90\columnwidth]{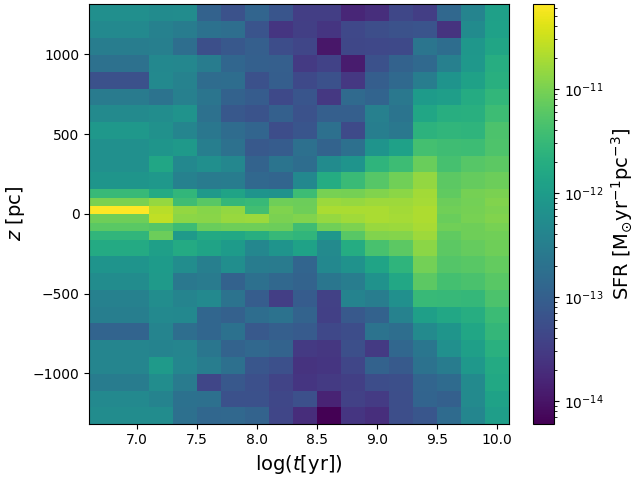}
    \caption{Plot of the SFR per unit volume, as determined for all slices and all age bins. It can be seen how the young stars are mostly concentrated on the Galactic Plane while older ones are more dispersed.}
    \label{fig:SFR_2D}
\end{figure}
Figure~\ref{fig:SFR_2D} illustrates the main result of this paper, namely the SFR per unit volume in every slice and for all age bins.
The information on the SFR is also presented in Table~\ref{tab:sfr_results}, together with the error bars. There are multiple ways in which this information can be interpreted.

\subsection{The SFR as a function of height \texorpdfstring{$z$}{z}}
\label{sec:sfrXheight}

From Fig.~\ref{fig:SFR_2D}, it is evident that the SFR strongly varies with $z$, and in a way which is almost symmetrical with respect to $z=0$. 

\begin{figure*}
    \centering
    \begin{minipage}[t]{0.90\columnwidth}
        \centering
        $0~\mathrm{pc}<z<52.63~\mathrm{pc}$\\
        \includegraphics[width=\textwidth, trim={0 0 0 30px}, clip]{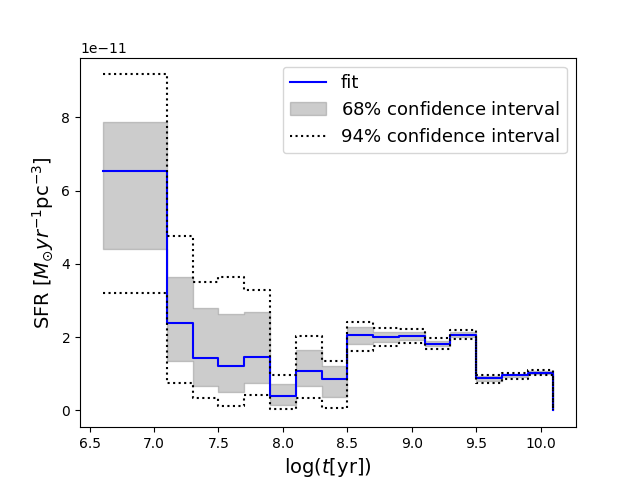}
    \end{minipage}
    \begin{minipage}[t]{0.90\columnwidth}
        \centering
        $368.42~\mathrm{pc}<z<5473.68~\mathrm{pc}$\\
        \includegraphics[width=\textwidth, trim={0 0 0 30px}, clip]{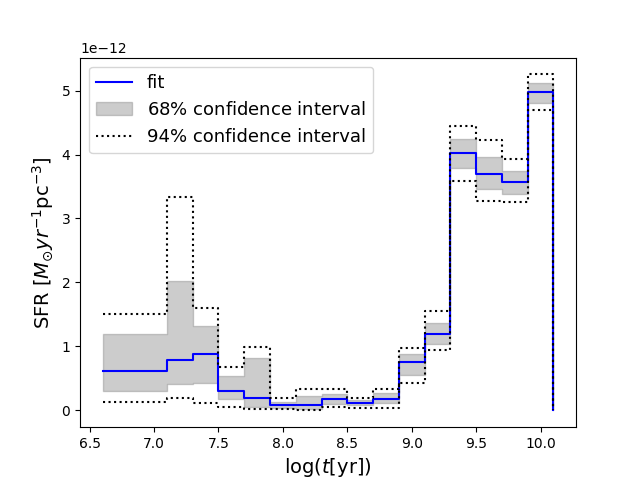}
    \end{minipage}\\
    \begin{minipage}[t]{0.90\columnwidth}
        \centering
        $684.21~\mathrm{pc}<z<789.47~\mathrm{pc}$\\
        \includegraphics[width=\textwidth, trim={0 0 0 30px}, clip]{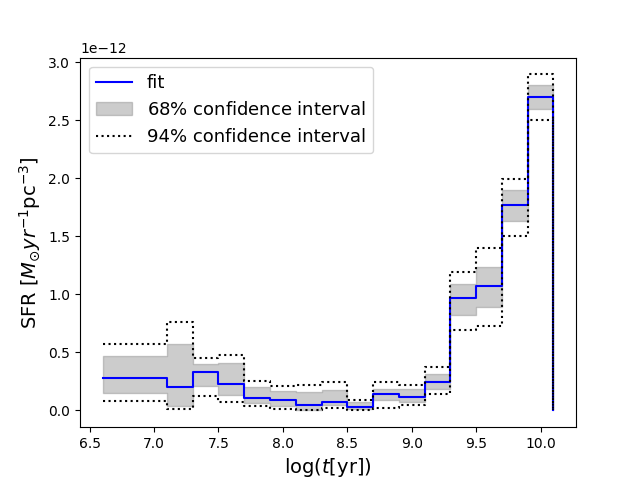}
    \end{minipage}
    \begin{minipage}[t]{0.90\columnwidth}
        \centering
        $894.74~\mathrm{pc}<z<1000.00~\mathrm{pc}$\\
        \includegraphics[width=\textwidth, trim={0 0 0 30px}, clip]{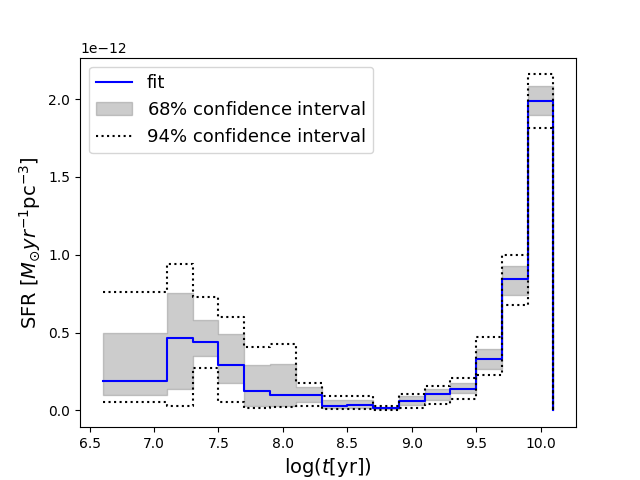}
    \end{minipage}\\
    \begin{minipage}[t]{0.90\columnwidth}
        \centering
        $1210.52~\mathrm{pc}<z<1315.78~\mathrm{pc}$\\
        \includegraphics[width=\textwidth, trim={0 0 0 30px}, clip]{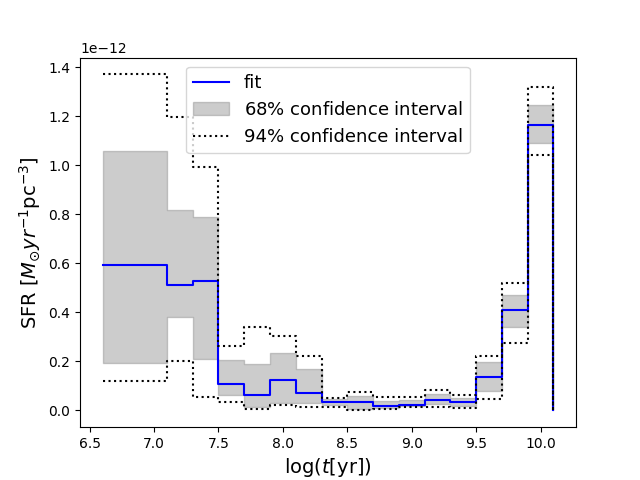}
    \end{minipage}
    \caption{The SFR per unit volume as a function of age determined for the same slices of the sample shown in Fig.~\ref{fig:results_comparison}, also indicated on top of each panel. The blue lines represent the best fitting SFR values, while the grey shaded areas and the dashed lines mark the $68\%$ and $94\%$ confidence intervals, respectively. We note that the SFR at young ages ($\logtyr\lesssim7.5$) is non-zero even at large $z$, although it is very uncertain.}
    \label{fig:result_sfr}
\end{figure*}

Figure~\ref{fig:result_sfr} presents the best-fit star formation rates (SFR) for the same slices as in Fig.~\ref{fig:results_comparison}. It can be appreciated that as we increase $z$ from $0$, the peak of SFR progressively moves towards older ages, and the overall SFR decreases.

An important clarification is necessary at this point: our method determines the SFR \textit{where it is observed now}, and not at their place of birth. With regard to the vertical distribution, it means that the old populations that are now observed at high $|z|$ could have (and probably have) been born closer to the Galactic Plane, and just got dispersed vertically as age increased (see next subsection). The same can be said with regard to the radial distribution in the Galaxy: older populations might have been born in different disc locations, and be observed now while at the Solar Neighbourhood (in a process commonly known as radial migration, see \citealt{lu22} and references therein). This uncertainty about the place of birth of stellar populations is common to all SFH determinations in nearby galaxies.

We can qualitatively compare our results close to the Galactic Plane with the results obtained by \cite{daltio21} for a \gaia DR2 sample within 200~pc of the Sun and excluding low galactic latitudes. For this comparison, we simply integrate the SFR obtained for all slices with $|z|<250$~pc.
We find the same peak of SFR in the $\logtyr=9.3-9.5$ age bin, however we do not recover the peaks that they found at other ages. The differences do not appear substantial, and might be attributed to the different samples and methods, and the better understanding of the \gaia errors in DR3. 

We can compare in the same way our results to those obtained by \cite{alzate20} for a sample of \gaia DR2 photometry within 100~pc from the Sun, by adding our best fitting SFR for slices with $|z|<105$~pc.
A comparison is not straightforward: their results are presented using the relative weight $a_{\mathrm{AGE},i}$ of each isochrone, we do not have enough age resolution to compare the results of the oldest SFR down to $t=4.8$~Gyr and they refer to slightly different galactocentric radii.
Nevertheless, in both cases the SFR appears to be declining from this age down to $t\approx 0.1$~Gyr, the limit of their age interval.
Finally, we find a peak in SFR at $t\approx 3$~Gyr that is not found in their results.

\subsection{The scale height as a function of age}
\label{sec:heightXage}
From the best fitting SFRs of each slice, we can compute the scale height of the disc at every age bin, $h_{z}(t)$, by fitting the expected profile of the disc for the SFRs. In general, two functions have been used to describe this profile: the classical exponential
\begin{equation} \label{eq:sfr_exp}
    f_{\exp}(t,z) = A \exp{\left(-|z|/h_{z}(t)\right)} + c
\end{equation}
that applies well to stars at large heights \citep[e.g.][]{gilmore83,juric08}, and the
square hyperbolic secant
\begin{equation} \label{eq:sfr_sech2}
    f_{\sechsq}(t,z) = A \sechsq{\left(-z/h_{z}(t)\right)} + c
\end{equation}
which should apply to a single-component, isothermal disc \citep{spitzer42,bahcall84}. None of these functions should be strictly valid in the case of a multi-component disc as the MW one \citep[see e.g.][]{sarkar20}; however, they are still useful approximations, valid at high $|z|$, and they allow an easy comparison with previous results in the literature. We note that we have also introduced a constant $c$ in both equations, which is usually absent.

\begin{figure*}
    \centering
    \begin{tabular}{cccc}
        \includegraphics[width=0.90\columnwidth]{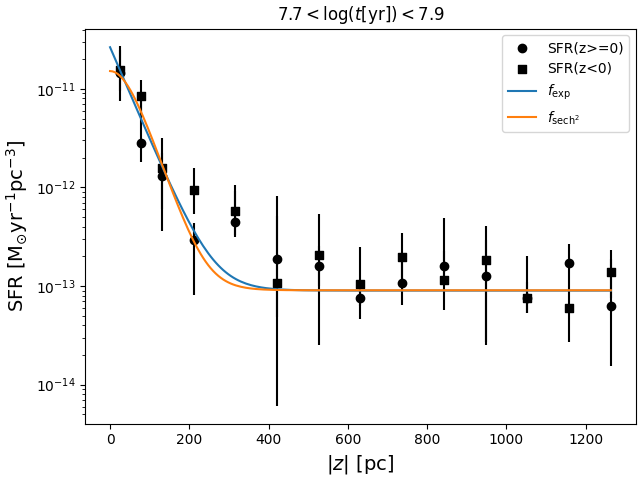} &
        \includegraphics[width=0.90\columnwidth]{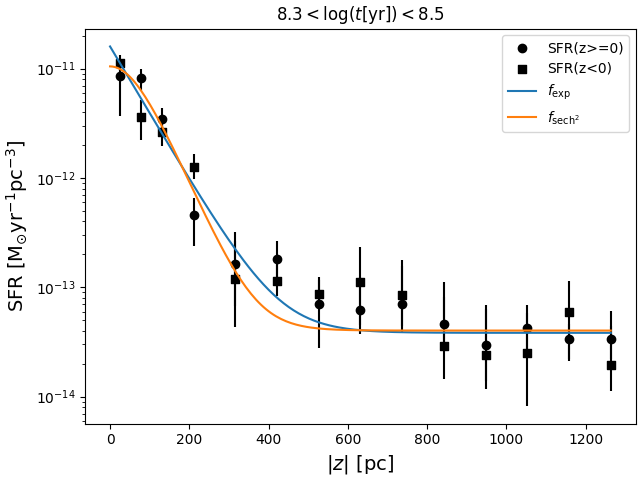}\\
        \includegraphics[width=0.90\columnwidth]{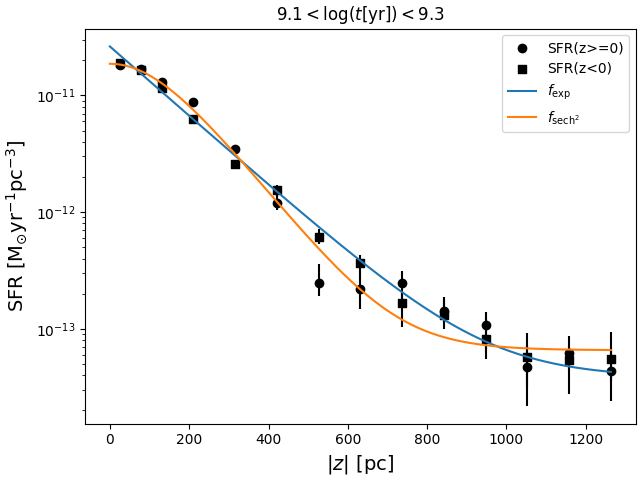} &
        \includegraphics[width=0.90\columnwidth]{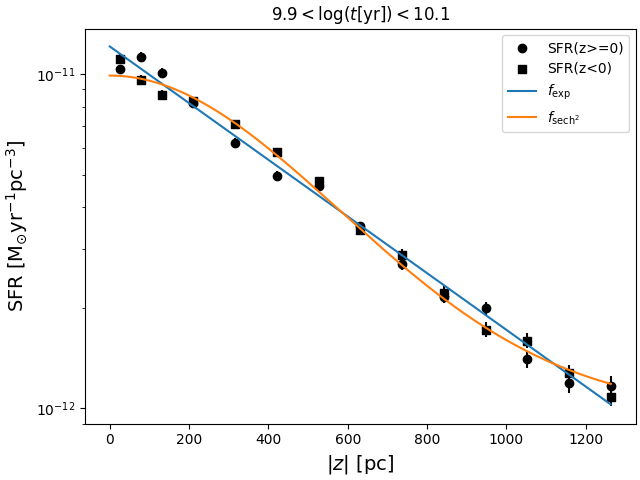}\\
    \end{tabular}
    \caption{Fit of the SFR per unit volume (black circles and squares) with the exponential (blue) and squared hyperbolic secant (orange) functions.}
    \label{fig:sfr_hz_fit}
    \centering
    \begin{tabular}{cc}
        \includegraphics[width=0.90\columnwidth]{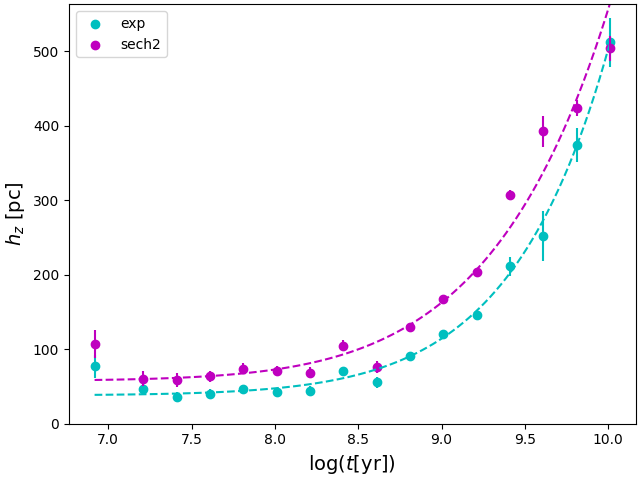} &
        \includegraphics[width=0.90\columnwidth]{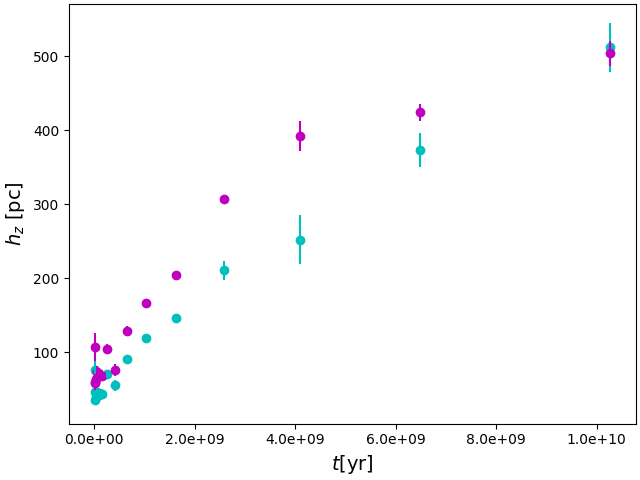}\\
    \end{tabular}
    \caption{The scale height of the Galactic Disc $h_{z}$ versus the mean age of every age bin. The cyan points are the exponential case, while the magenta points are for the case of the square hyperbolic secant. The left plot uses a logarithmic scale for the age, while the right one uses a linear one and shows that the linear trend in younger ages is lost at older ages, where the trend in $h_{z}$ flattens. The dashed lines in the left panel show the results of the fits of Eq.~\ref{eq:hzt}.}
    \label{fig:hz_vs_age}
\end{figure*}

We use both functional forms to fit the trend of SFR with $|z|$ and examples of these fits performed in different ages bins are presented in Fig.~\ref{fig:sfr_hz_fit}.
This figure clearly illustrates the difficulty in selecting a single functional form to describe the SFR profile with $|z|$ at all ages.
While achieving the optimal model selection is not in the scope of this work, it can be appreciated how the result in the $9.1<\logtyr<9.3$ age bin appears to be better fitted by the squared hyperbolic secant, at least for $|z|<600$~pc, while the one in the $9.9<\logtyr<10.1$ bin appears to favour an exponential fit. Another aspect that becomes clear especially in the upper (younger) two panels, is the need for the constant $c$ in Eq.~\ref{eq:sfr_exp} and \ref{eq:sfr_sech2} for the younger ages, the lack of which would not allow a proper fit with either function of the flatter trend of the SFR at large $|z|$.

The overall result of the SFR fits is summarised in Table~\ref{tab:result_hz} and in Fig.~\ref{fig:hz_vs_age}, and it clearly shows the tendency of $h_{z}$ to increase with age, irrespective of the functional form used.
We note that there are two important details in this figure:
\begin{itemize}
    \item The scale heights for all young ages ($t\lesssim 500$~Myr) are essentially constant with values smaller than $100$~pc. However, the scale heights for the age bin with $7.7<\logtyr<7.9$ are notably higher, particularly in the youngest age bin.
    \item For all ages smaller than $4$~Gyr, the scale height appears to have an approximately linear trend with age. For the $\sechsq$ case, the $h_{z}$ flattens at values close to $450$~pc for all older ages. For the $\exp$ case, some flattening occurs at later ages, but it is not so marked.
\end{itemize}

\begin{table*}
    \caption{Galactic Disc scale height resulting from fitting Eq.~\ref{eq:sfr_exp} and Eq.~\ref{eq:sfr_sech2} to the fitted SFRs.}
    \label{tab:result_hz}
    \centering
    \begin{adjustbox}{center}
    \setlength\tabcolsep{1.0pt}
    \begin{tabular}{@{\extracolsep{2pt}}ccccccccccc@{}}
        \toprule
         & & \multicolumn{4}{c}{$f_{\exp}$} & \multicolumn{4}{c}{$f_{\sechsq}$} \\
        \cline{3-6}
        \cline{7-10}
        $\logtyri{1}$ & $\logtyri{2}$
        & $A$ & $h_{z}$ & $c$ & $\chi^{2}$
        & $A$ & $h_{z}$ & $c$ & $\chi^{2}$ \\
         &  
        & $[10^{-11}\Msun\mathrm{yr}^{-1}\mathrm{pc}^{-3}]$ & [pc] & $[10^{-11}\Msun\mathrm{yr}^{-1}\mathrm{pc}^{-3}]$ &
        & $[10^{-11}\Msun\mathrm{yr}^{-1}\mathrm{pc}^{-3}]$ & [pc] & $[10^{-11}\Msun\mathrm{yr}^{-1}\mathrm{pc}^{-3}]$ & \\
        \midrule
        6.60 & 7.10  & 1.695 $\pm$ 0.674 &  76.7 $\pm$ 15.6 & 0.02605 $\pm$ 0.00462 &  25.8 & 1.178 $\pm$ 0.445 & 106.9 $\pm$ 19.2 & 0.26945 $\pm$ 0.00465 &  27.1 \\
        7.10 & 7.30  & 3.860 $\pm$ 1.244 &  46.8 $\pm$  8.6 & 0.04999 $\pm$ 0.00531 &  13.3 & 2.858 $\pm$ 0.816 &  60.3 $\pm$  9.8 & 0.50915 $\pm$ 0.00539 &  14.0 \\
        7.30 & 7.50  & 2.797 $\pm$ 1.149 &  36.2 $\pm$  6.6 & 0.03326 $\pm$ 0.00292 &  12.8 & 1.497 $\pm$ 0.569 &  58.2 $\pm$  9.0 & 0.33437 $\pm$ 0.00292 &  13.0 \\
        7.50 & 7.70  & 3.040 $\pm$ 1.215 &  40.1 $\pm$  6.3 & 0.01333 $\pm$ 0.00218 &  19.7 & 1.798 $\pm$ 0.611 &  63.4 $\pm$  7.6 & 0.13443 $\pm$ 0.00212 &  18.7 \\
        7.70 & 7.90  & 2.637 $\pm$ 0.743 &  46.3 $\pm$  5.3 & 0.00899 $\pm$ 0.00099 &  11.4 & 1.507 $\pm$ 0.409 &  73.7 $\pm$  7.6 & 0.00906 $\pm$ 0.00101 &  11.9 \\
        7.90 & 8.10  & 1.769 $\pm$ 0.565 &  41.8 $\pm$  5.3 & 0.00784 $\pm$ 0.00119 &  28.6 & 1.071 $\pm$ 0.275 &  70.3 $\pm$  6.6 & 0.00788 $\pm$ 0.00111 &  25.0 \\
        8.10 & 8.30  & 2.245 $\pm$ 0.704 &  43.9 $\pm$  6.1 & 0.00513 $\pm$ 0.00078 &  33.7 & 1.427 $\pm$ 0.374 &  68.4 $\pm$  7.2 & 0.00516 $\pm$ 0.00075 &  31.1 \\
        8.30 & 8.50  & 1.590 $\pm$ 0.221 &  70.9 $\pm$  4.9 & 0.00383 $\pm$ 0.00055 &  20.9 & 1.043 $\pm$ 0.144 & 104.8 $\pm$  6.9 & 0.00402 $\pm$ 0.00057 &  22.9 \\
        8.50 & 8.70  & 2.822 $\pm$ 0.657 &  55.3 $\pm$  7.7 & 0.00226 $\pm$ 0.00113 & 176.4 & 2.018 $\pm$ 0.401 &  76.1 $\pm$  8.4 & 0.00234 $\pm$ 0.00113 & 175.0 \\
        8.70 & 8.90  & 2.625 $\pm$ 0.203 &  90.7 $\pm$  4.7 & 0.00270 $\pm$ 0.00074 &  72.2 & 1.816 $\pm$ 0.134 & 129.2 $\pm$  6.0 & 0.00297 $\pm$ 0.00075 &  75.1 \\
        8.90 & 9.10  & 2.804 $\pm$ 0.143 & 120.0 $\pm$  4.4 & 0.00315 $\pm$ 0.00108 &  80.9 & 1.974 $\pm$ 0.090 & 167.0 $\pm$  5.1 & 0.00412 $\pm$ 0.00099 &  71.9 \\
        9.10 & 9.30  & 2.625 $\pm$ 0.147 & 145.8 $\pm$  6.3 & 0.00382 $\pm$ 0.00222 & 179.6 & 1.862 $\pm$ 0.064 & 204.0 $\pm$  5.0 & 0.00655 $\pm$ 0.00135 &  76.6 \\
        9.30 & 9.50  & 2.972 $\pm$ 0.196 & 211.1 $\pm$ 12.8 & 0.00000 $\pm$ 0.00653 & 573.8 & 2.096 $\pm$ 0.053 & 307.1 $\pm$  6.0 & 0.00163 $\pm$ 0.00218 & 102.3 \\
        9.50 & 9.70  & 1.230 $\pm$ 0.135 & 252.2 $\pm$ 33.5 & 0.00000 $\pm$ 0.00814 & 396.6 & 0.865 $\pm$ 0.048 & 392.4 $\pm$ 20.4 & 0.00000 $\pm$ 0.00360 & 138.3 \\
        9.70 & 9.90  & 1.170 $\pm$ 0.040 & 373.7 $\pm$ 23.0 & 0.00000 $\pm$ 0.01244 &  92.1 & 0.890 $\pm$ 0.021 & 424.0 $\pm$ 11.1 & 0.03903 $\pm$ 0.00494 &  48.7 \\
        9.90 & 10.10 & 1.210 $\pm$ 0.027 & 512.0 $\pm$ 32.6 & 0.00000 $\pm$ 0.02142 & 104.8 & 0.896 $\pm$ 0.023 & 503.7 $\pm$ 17.3 & 0.09480 $\pm$ 0.01016 & 125.9 \\
        \bottomrule
    \end{tabular}
    \end{adjustbox}
\end{table*}

We remark that the constant $c$ has been introduced in our fits because it improves a lot the fitting of both equations, especially in the case of young age bins.
As can be seen in Table~\ref{tab:result_hz}, it becomes completely irrelevant (i.e., compatible with a $c=0$ value) for all ages $\log t[\mathrm{yr}]>8.5$ in both the exponential case and the $\sechsq$ cases, with the exception of the $\sechsq$ fit at the two oldest ages.
For young ages, instead, the value of this constant is sizeable, but does not exceed 1.5 per cent in the former and 2.1 per cent in the latter of the value of the constant $A$, that is the value of the functions at $z=0$.
This constant is needed because, at all young ages, there is always a ``background'' of star formation being detected at all $z$, even very far from the Galactic Plane (see Figs~\ref{fig:result_sfr} and \ref{fig:results_comparison}).
This background does not disappear with simple changes in the fitting method (see Appendix~\ref{sec:young_sfr} of the online supplementary material). At least partially, it reflects the presence  of real star counts along the bright main sequence at all $z$, as can be appreciated by the stars with $\gmagabs<2$ and $\gcolorabs<0.5$ in the observations of Fig.~\ref{fig:results_comparison}.
It is not evident whether these main sequence stars are truly young.
They could also be blue stragglers, which our binary models predict but maybe not in the right numbers to fit the observations -- hence leading to the fitting of those stars by means of a young star formation.
To clarify this feature, we need first to examine in detail the predictions of the binary BinaPSE module by means of special samples, which is not straightforward, and is not among the aims of this paper.

It should be noted at this point that this kind of determination of the scale height $h_{z}$ with high age resolution is currently possible only for the case of the Milky Way. For external galaxies, there are alternative opportunities, given the availability of kinematics \citep[see e.g.][]{dorman15} and methods based on the dust geometry in inclined discs \citep[e.g.][]{dalcanton23}. However, the age resolution attainable in such cases is quite limited.

Finally, we attempt to derive a fitting formula for the relation between $h_{z}$ and age, using the function 
\begin{equation}
    h_{z}(t) = h_{z,0} \left(1 + \frac{t}{\tau}\right)^{n}
    \label{eq:hzt}
\end{equation}
\citep[eq.~26 of][]{villumsen83}, where $h_{z,0}$ is the scale height at $t=0$, $\tau$ is a timescale for the evolution of $h_{z}$ and $n$ controls the steepness of the relation.
Fitting the values of $h_{z}$ in Table~\ref{tab:result_hz} we find
\begin{equation}
    \begin{split}
        h_{z,0} &= 37.1 \pm 4.8~\mathrm{pc}\\
        \tau &= (2.7 \pm 1.0) \times 10^{8}~\mathrm{yr}\\
        n &= 0.714 \pm 0.061
    \end{split}
\end{equation}
for Eq.~\ref{eq:sfr_exp} and
\begin{equation}
    \begin{split}
        h_{z,0} &= 55.8 \pm 8.4~\mathrm{pc}\\
        \tau &= (1.77 \pm 0.95) \times 10^{8}~\mathrm{yr}\\
        n &= 0.566 \pm 0.063
    \end{split}
\end{equation}
for Eq.~\ref{eq:sfr_sech2}, and the left panel of Fig.~\ref{fig:hz_vs_age} shows the two fits as dashed lines.
Our best fitting $n$ in the first case is remarkably close to the value of $2/3$ favoured by \citet[][eq.~4]{rana92}, but the best fitting $\tau$ is about half of their value; the second case has a value of $n$ closer to $1/2$, and an even smaller $\tau$.

We note that Eq.~\ref{eq:hzt} is based on a simple model for the scattering of disc stars by giant molecular clouds \citep{villumsen83}, and subsequent fits of the increase in velocity dispersion with age in their simulations \citep{villumsen85,rana92}. This model cannot be considered as a reference in modern times, given that it ignores important dynamical processes such as radial migration \citep{schonrich09} and the early accretion of small galaxies in the Galactic disc \citep{helmi20}. However, it is remarkable that the simple form of Eq.~\ref{eq:hzt} is still useful to represent the actual measurements of $h_{z}$, now derived independently from any kinematical model or data.
 
\subsection{Comparison with scale heights derived from simple star counts}
\begin{figure}
    \centering
    \includegraphics[width=0.90\columnwidth]{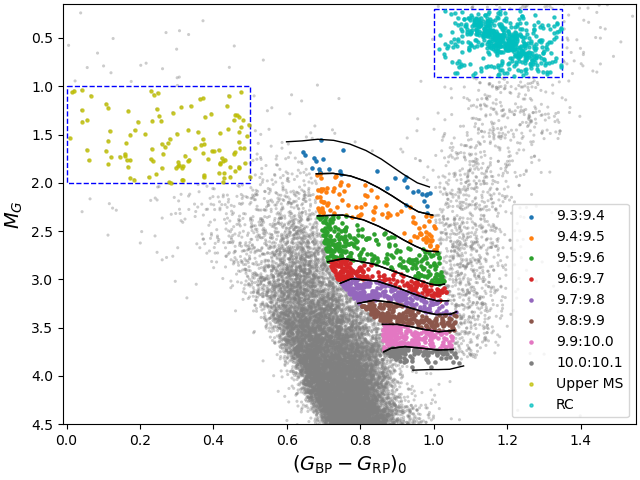}
    \caption{Regions of the CAMD used to determine the star counts for the slice with $-368.42~\mathrm{pc} < z < -263.16$~pc. The isochrones correspond to the SGB phase at different ages (black lines, from $\logtyr=9.3$ to $\logtyr=10.1$) and the coloured points between them are the sources we selected for each age interval. The dashed boxes instead select young Main Sequence (left, yellow points) and Red Clump (right, cyan points) stars.}
    \label{fig:counts_cmd}
\end{figure}

\begin{figure}
    \centering
    \includegraphics[width=0.90\columnwidth]{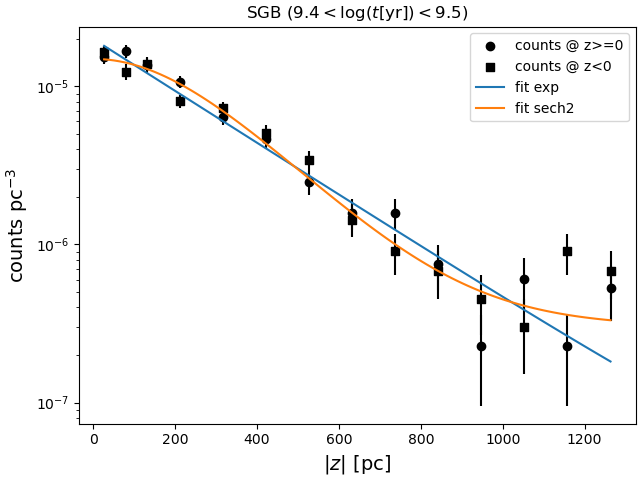}\\
    \includegraphics[width=0.90\columnwidth]{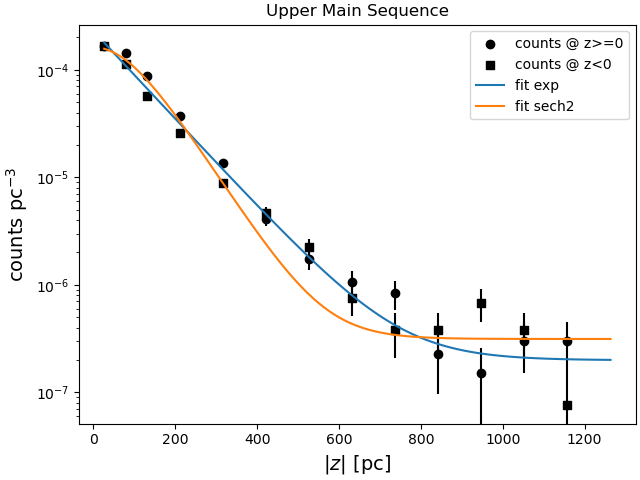}\\
    \includegraphics[width=0.90\columnwidth]{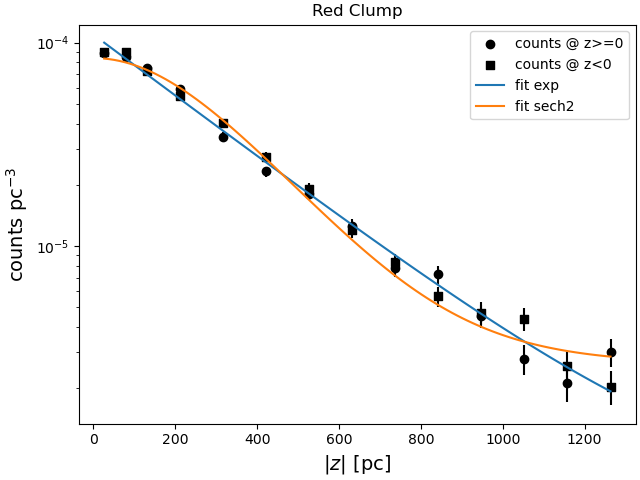}
    \caption{Star counts per unit volume from the data and their fits with the $\exp$ (blue lines) and $\sechsq$ (orange lines) functions. The shape of the black markers indicate counts at $z\geq0$ (circles) and $z<0$ (squares).
    Panels from the top to the bottom correspond to stars selected in three different regions illustrated in Fig.~\ref{fig:counts_cmd}: SGB stars between the isochrones of $\logtyr=9.4$ and $\logtyr=9.5$, main sequence stars, and red clump stars, respectively.}
    \label{fig:counts}
\end{figure}
\begin{table*}
    \caption{Best fit parameters for the exponential ($h_{z,\exp}$) and squared hyperbolic secant ($h_{z, \sechsq}$) fits. The first column specifies the region of the CAMD used: the first block of rows refers to star counts between isochrones with the given ages at the sub-giant branch (SGB) phase (upper panel of Fig.~\ref{fig:counts}), while the second block uses simple CAMD ``boxes'' (middle and bottom panel of Fig.~\ref{fig:counts}). The second and third columns, in the case of the SGB, show the ages of the two isochrones between which the star counts are computed.}
    \label{tab:counts_zlice1_fit}
    \makebox[\textwidth]{%
        \setlength\tabcolsep{1.0pt}
        \begin{tabular}{@{\extracolsep{1.5pt}}ccccccccccc@{}}
            \toprule
            & & & \multicolumn{4}{c}{$f_{\exp}$} & \multicolumn{4}{c}{$f_{\sechsq}$} \\
            \cline{4-7}
            \cline{8-11}
            Phase & $\logtyri{1}$ & $\logtyri{2}$
            & $A$ & $h_{z}$ & $c$ & $\chi^{2}$
            & $A$ & $h_{z}$ & $c$ & $\chi^{2}$\\
             & & 
            & $[10^{-5}\text{counts}\,\mathrm{pc}^{-3}]$ & [pc] & $[10^{-5}\text{counts}\,\mathrm{pc}^{-3}]$ &
            & $[10^{-5}\text{counts}\,\mathrm{pc}^{-3}]$ & [pc] & $[10^{-5}\text{counts}\,\mathrm{pc}^{-3}]$ & \\
            \midrule
            \multirow{8}{*}{SGB} &  9.3 &  9.4 &  0.800 $\pm$ 0.062 & 211.4 $\pm$ 16.8 & 0.0080 $\pm$ 0.0048 & 25.4 & 0.577 $\pm$ 0.040 & 278.6 $\pm$ 16.1 & 0.0144 $\pm$ 0.0038 & 24.1 \\
                                 &  9.4 &  9.5 &  1.990 $\pm$ 0.123 & 264.2 $\pm$ 19.5 & 0.0015 $\pm$ 0.0146 & 52.5 & 1.461 $\pm$ 0.071 & 335.3 $\pm$ 14.5 & 0.0301 $\pm$ 0.0084 & 37.7 \\
                                 &  9.5 &  9.6 &  5.009 $\pm$ 0.221 & 296.3 $\pm$ 16.8 & 0.0000 $\pm$ 0.0319 & 75.0 & 3.693 $\pm$ 0.104 & 376.8 $\pm$  9.8 & 0.0685 $\pm$ 0.0148 & 36.2 \\
                                 &  9.6 &  9.7 &  3.623 $\pm$ 0.154 & 318.7 $\pm$ 19.4 & 0.0000 $\pm$ 0.0283 & 53.3 & 2.646 $\pm$ 0.076 & 408.7 $\pm$ 11.7 & 0.0497 $\pm$ 0.0137 & 29.2 \\
                                 &  9.7 &  9.8 &  3.378 $\pm$ 0.149 & 388.1 $\pm$ 31.6 & 0.0000 $\pm$ 0.0497 & 64.8 & 2.536 $\pm$ 0.078 & 463.3 $\pm$ 16.1 & 0.0832 $\pm$ 0.0217 & 35.5 \\
                                 &  9.8 &  9.9 &  3.126 $\pm$ 0.107 & 445.3 $\pm$ 36.1 & 0.0000 $\pm$ 0.0585 & 39.8 & 2.352 $\pm$ 0.077 & 470.2 $\pm$ 19.2 & 0.1689 $\pm$ 0.0270 & 35.4 \\
                                 &  9.9 & 10.0 &  3.096 $\pm$ 0.114 & 542.8 $\pm$ 63.6 & 0.0000 $\pm$ 0.1062 & 50.5 & 2.285 $\pm$ 0.090 & 576.5 $\pm$ 35.8 & 0.1804 $\pm$ 0.0566 & 57.5 \\
                                 & 10.0 & 10.1 &  1.584 $\pm$ 0.116 & 367.7 $\pm$ 47.0 & 0.0000 $\pm$ 0.0320 & 77.0 & 1.109 $\pm$ 0.076 & 497.9 $\pm$ 41.9 & 0.0117 $\pm$ 0.0239 & 84.8 \\
            \midrule
            \makecell{Upper main\\sequence} & & & 22.892 $\pm$ 1.160 & 106.3 $\pm$  4.0 & 0.0199 $\pm$ 0.0115 &  167.4 & 16.184 $\pm$ 0.882 & 147.0 $\pm$ 5.5 & 0.0314 $\pm$ 0.0127 & 219.5\\
            Red Clump                       & & & 10.890 $\pm$ 0.307 & 288.3 $\pm$ 10.2 & 0.0571 $\pm$ 0.0436 & 61.8 &  8.131 $\pm$ 0.257 & 347.0 $\pm$ 9.9 & 0.2639 $\pm$ 0.0341 & 87.6 \\
            \bottomrule
        \end{tabular}
    }
\end{table*}
\begin{figure}
    \centering
    \includegraphics[width=0.90\columnwidth]{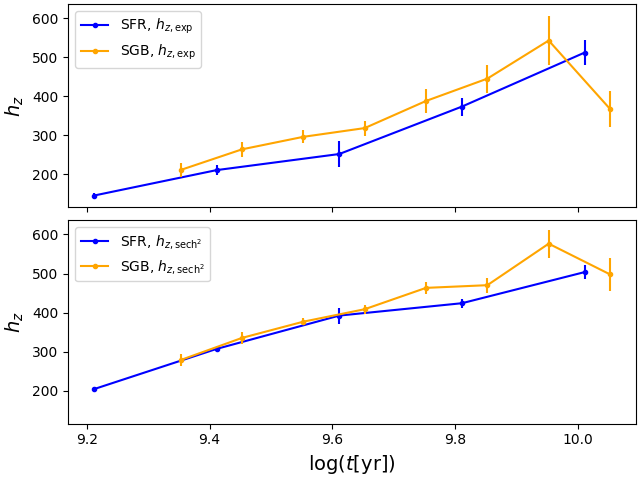}
    \caption{Comparison of $h_{z}$ obtained from the fits of the SFR and from the counts on the SGB, limited to the ages available in the latter case. \textit{Upper panel}. Exponential fit. \textit{Lower panel}. Squared hyperbolic secant fit.}
    \label{fig:compare_hz_SGB}
\end{figure}

A traditional way of deriving structural parameters of the MW disc is to determine the star counts of a specific kind of star for which distances can be derived, and fit them as a function of distance with exponentials. The method provides robust numerical results when applied to abundant objects such as Red Clump \citep[RC;][]{cabrera07} or Main Sequence stars \citep[MS;][]{juric08,bovy17,ferguson17}, but suffers from one major problem: as the objects that are abundant generally sample a quite large interval of stellar ages, only an age-averaged scale height can be derived. RC stars, for instance, span the entire age interval from 1.5 to 10 Gyr, with some concentration around the younger ages \citep{girardi16}, while the lower MS ones evenly sample all ages larger than a few Myr.

With the \gaia catalogue, this problem can be circumvented, to some extent, by using stars along the subgiant branch (SGB), which separate reasonably well in age. In the following, we attempt to derive the vertical scale height of the Disc at different ages using their star counts, for the sake of providing a useful comparison with the results we have obtained in Sect.~\ref{sec:heightXage}.

We take advantage of PARSEC v1.2S solar-metallicity isochrones \citep{bressan12} computed at ages $9.3 \leq \logtyr) \leq 10.3$, with a step of $0.1$~dex in $\logtyr$, then select all the observed stars between the SGB portions of two subsequent isochrones, as illustrated in Fig.~\ref{fig:counts_cmd}.
To compare with the results obtained without age resolution, we also select stars in two other regions of the CAMD, namely the upper MS (which includes stars with $7.6 \lesssim \logtyr \lesssim 9.2$) and the RC, using simple CAMD boxes.
Table~\ref{tab:counts_zlice1} shows the counts we derive for all slices and age ranges, and the fit for a single age interval on the SGB and for the two CAMD boxes are shown in Fig.~\ref{fig:counts}. Finally, Table~\ref{tab:counts_zlice1_fit} summarises the scale heights we obtain.

Visual inspection shows that most fits are of good quality.
In the case of SGB star counts, the resulting scale heights are compatible with the values derived from the SFR for the same age bins (Table~\ref{tab:result_hz} and Fig.~\ref{fig:compare_hz_SGB}), generally confirming our previous conclusions about the increase of $h_z$ with age. This appears to confirm the suitability of the simpler method based on counts of SGB stars. The same cannot be said for the scale heights derived from the CAMD boxes. Although the fitting of RC star counts is also excellent (bottom panel of Fig.~\ref{fig:counts}), $h_{z}\sim300-350$~pc clearly corresponds to an intermediate value, valid at ages of $\sim3$ Gyr, and cannot be assigned to the entire disc. Similarly, for the upper MS $h_{z}\sim100-150$~pc appears to be more in line with the scale height of the older part ($t \gtrsim 9$~Gyr) of its age interval.

More generally, we note that there is no single place in the CAMD which corresponds exactly to a single-burst stellar population, especially when we consider the populations of binaries, and the possibility of having a broad age-metallicity relation at old ages. As a result, scale heights derived from simple star counts will always be somewhat smeared out with respect to the true ones, even when referred to the SGB sections. Moreover, the SGB method does not apply to young ages. 

Our method based on the fitted SFR represents an attempt to reduce these uncertainties. At the same time, it directly provides a quantitative measure of the total mass of stars involved in every fitted exponential or squared hyperbolic secant function, therefore telling us how different discs compare in terms of total mass or surface density, as we discuss in the following subsection.

\subsection{The vertically-integrated SFR}
\begin{figure}
    \centering
    \includegraphics[width=0.90\columnwidth]{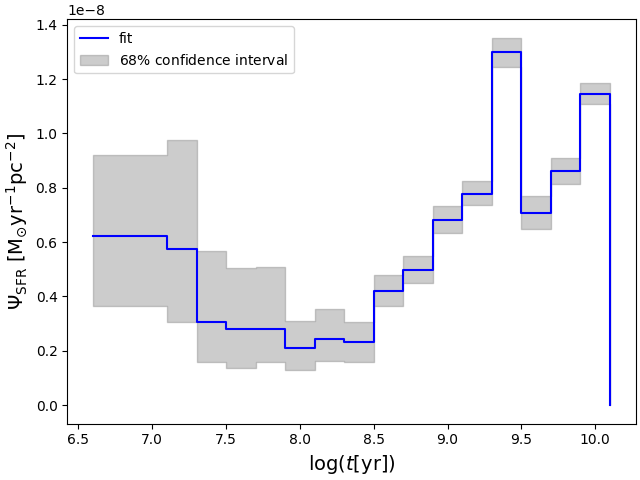}
    \caption{Overall surface SFR density of the Solar Cylinder at each age bin.}
    \label{fig:sfr_tot}
\end{figure}

\begin{table}
    \centering
    \caption{Total SFR per unit area for each age bin, computed in 3 different ways: just adding the fitted SFR (third column), and integrating the exponential and square hyperbolic secant functions (fourth and fifth columns, respectively) between $z=-1315.78~\text{pc}$ and $z=1315.78~\text{pc}$.}
    \label{tab:surface_sfr}
    \begin{adjustbox}{center}
        \begin{tabular}{ccccc}
            \toprule
            $\logtyri{1}$ & $\logtyri{2}$
            & $\Psi_{\mathrm{SFR}}$
            & $\Psi_\mathrm{\exp}(t)$
            & $\Psi_\mathrm{\sechsq}(t)$ \\
            \cline{3-5}
             &  
            & \multicolumn{3}{c}{$[10^{-9}\Msun\mathrm{yr}^{-1}\mathrm{pc}^{-2}]$}\\
            \midrule
            6.60 & 7.10  & $ 6.20_{-2.60}^{+3.00}$ &  3.29 &  3.14 \\[2pt]
            7.10 & 7.30  & $ 5.70_{-2.70}^{+4.00}$ &  4.93 &  4.63 \\[2pt]
            7.30 & 7.50  & $ 3.10_{-1.50}^{+2.60}$ &  2.90 &  2.52 \\[2pt]
            7.50 & 7.70  & $ 2.80_{-1.40}^{+2.20}$ &  2.79 &  2.59 \\[2pt]
            7.70 & 7.90  & $ 2.80_{-1.20}^{+2.30}$ &  2.68 &  2.43 \\[2pt]
            7.90 & 8.10  & $ 2.11_{-0.82}^{+1.00}$ &  1.69 &  1.69 \\[2pt]
            8.10 & 8.30  & $ 2.44_{-0.83}^{+1.09}$ &  2.11 &  2.07 \\[2pt]
            8.30 & 8.50  & $ 2.31_{-0.72}^{+0.76}$ &  2.36 &  2.28 \\[2pt]
            8.50 & 8.70  & $ 4.21_{-0.54}^{+0.59}$ &  3.18 &  3.12 \\[2pt]
            8.70 & 8.90  & $ 4.97_{-0.47}^{+0.54}$ &  4.83 &  4.76 \\[2pt]
            8.90 & 9.10  & $ 6.81_{-0.47}^{+0.50}$ &  6.81 &  6.69 \\[2pt]
            9.10 & 9.30  & $ 7.79_{-0.43}^{+0.45}$ &  7.76 &  7.75 \\[2pt]
            9.30 & 9.50  & $12.99_{-0.54}^{+0.53}$ & 12.52 & 12.89 \\[2pt]
            9.50 & 9.70  & $ 7.09_{-0.59}^{+0.59}$ &  6.17 &  6.74 \\[2pt]
            9.70 & 9.90  & $ 8.60_{-0.48}^{+0.49}$ &  8.49 &  8.37 \\[2pt]
            9.90 & 10.10 & $11.47_{-0.38}^{+0.39}$ & 11.44 & 11.00 \\[2pt]
            \bottomrule
        \end{tabular}
    \end{adjustbox}
\end{table}
From the results we get for each slice, we can compute the overall, vertically integrated SFR of the Solar Cylinder:
\begin{equation}
    \Psi_\mathrm{SFR}(t) = \sum_{n=1}^{N_\mathrm{slice}} \mathrm{SFR}_n(t) \, \Delta z_n  \, ,
\end{equation}
where $\Delta z_n = \zmax-\zmin$.
This function is presented in Fig.~\ref{fig:sfr_tot} and in Table~\ref{tab:surface_sfr}.
There appears to be an almost linearly decreasing surface SFR from the oldest ages up to the $\logtyr=8.3-8.5$ bin, broken in two phases by the peak in SFR at the $\logtyr=9.3-9.5$ bin.
At bins younger than this one, the surface SFR appears to be somewhat constant down to the $\logtyr=7.3-7.5$ bin.
The youngest age bins show an increase in SFR, but the large errors associated to these estimates prevent a definitive conclusion.

Similar quantities can be derived by integrating eqs.~\ref{eq:sfr_exp} and \ref{eq:sfr_sech2} from $z=-1315.78$~pc to $1315.78$~pc, giving origin to the functions $\Psi_\mathrm{\exp}(t)$ and $\Psi_\mathrm{\sechsq}(t)$. Their values are reported in Table~\ref{tab:surface_sfr}, together with $\Psi_\mathrm{SFR}(t)$. 
As can be noticed, the three estimates are very similar, except at very young ages.
The benefit of defining these latter quantities is that they enable us to easily implement our current results in population synthesis codes like TRILEGAL \citep{girardi05}. As a result, we can perform more accurate simulations of the Galactic Disc for various spatial samples and photometric systems.

Our vertically-integrated SFRs, $\Psi(t)$, represent a volume large enough to allow comparisons with other SFRs present in the literature. 
For instance, \cite{ruiz22} modelled \gaia DR2 data for a sphere of radius $2$~kpc centred on the Sun, finding three SFR peaks at ages 5.7, 1.9 and 1.0~Gyr. Direct comparisons with our results are not straightforward given the different age resolution adopted, but comparing their figure~2 with our Fig.~\ref{fig:sfr_tot} we can see that the overall trend of SFR constantly decreasing with age is similar. However, they find a strong peak in SFR around 6~Gyr, while we find one in the younger $9.3 < \logtyr < 9.5$ age bin (ages from 2 to 3 Gyr). Concerning the other SFR peaks, we likely do not have a good enough age resolution to see them. Interestingly, the results might agree at the youngest ages (younger than 0.1 Gyr), where in both cases the SFR appears to have a boost. 

Note that our 2-3 Gyr peak coincides with the one found by \citet{mor19}, and might correspond to the 3~Gyr peak introduced by \citet{syso22}. Both works are based on the modelling of the entire Solar Neighbourhood as sampled by Gaia.

\begin{figure}
    \includegraphics[width=0.90\columnwidth]{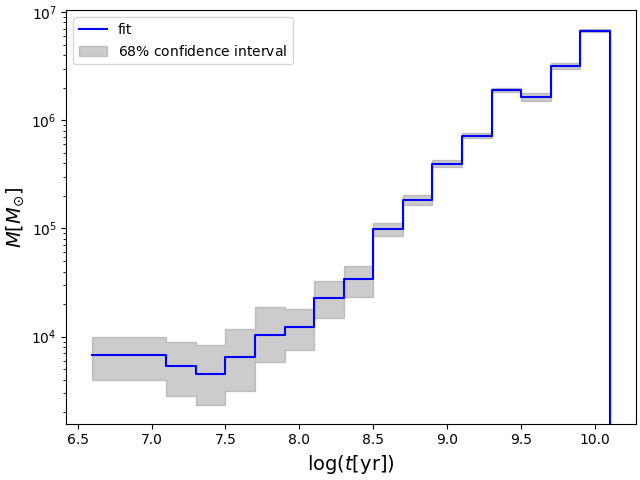}
    \caption{Total mass of stars produced in the Solar Cylinder in each age bin.}
    \label{fig:mass_produced}
\end{figure}

\subsection{The stellar mass formed in the Solar Cylinder}
The contribution of each age bin to the total mass of stars ever produced is shown in Fig.~\ref{fig:mass_produced}. Similarly to Fig.~\ref{fig:sfr_tot}, there is a decreasing amount of mass produced from the older to the younger ages, with a bump at the $\logtyr=9.3-9.5$ bin. Due to the increased SFR of the youngest ages, the trend appears to become constant for ages younger than the $\logtyr=7.5-7.7$ age bin. Nevertheless, the overall contribution of these young age bins to the total mass is minor compared to the other age bins.

Summing together all contributions over the interval of age we explored, we determine the total mass of stars ever formed in the Solar Cylinder:
\begin{equation}
    M_{\mathrm{tot}} = 1.492 \pm 0.079 \times 10^{7} \, \Msun
\end{equation}
and the surface density of stars ever formed:
\begin{equation}
    \Sigma_{\text{SFR}} = 118.7 \pm 6.2 \, \Msun \, \mathrm{pc}^{-2}
    \label{eq:surfacedensity}
\end{equation}
This quantity is of interest as it allows comparing our Galaxy with other disc galaxies for which the SFR cannot be investigated in the full 3D space, but their SFR density can be derived across the disc either by using resolved photometry from HST and JWST \citep[e.g.,][]{williams17,lazzarini22}, or integral field spectroscopy \citep[e.g.,][]{garcia17,lopez18}. In addition, from our results, we can also derive the contribution of each age bin to the surface SFR, which is shown in Table~\ref{tab:result_hz}.

We recall that a fraction of the mass we derive has been already returned to the ISM, and may have been locked up again in successive generations of stars, in amounts that cannot be quantified unless we build complex chemical evolution models of the MW disc. Therefore, the surface density of Eq.~\ref{eq:surfacedensity} is not expected to coincide with the present mass density in the Solar Neighbourhood, which is usually derived with the help of the stellar kinematics. It is nonetheless comforting that the total mass density of stars ever formed has the same order of magnitude as the kinematically-derived one, which amounts to $\Sigma_\mathrm{kin}(R_\odot) = 61-84\,\Msun\,\mathrm{pc}^{-2}$ at heights of $\sim1$~kpc \citep[e.g.][]{kuijken91, holmberg04,bienayme14,binney23}.

Moreover, we note that our estimates of $\Sigma_\mathrm{SFR}$ and $M_{\mathrm{tot}}$ could still be reduced, systematically, by factors of about 40 to 60 per cent, if the binary fraction that characterises our sample should turn out significantly smaller that the fitted value of $\fbin=0.984$. This aspect is discussed in App.~\ref{app:binary_uncert}.

\subsection{The apparent vertical metallicity gradient}

One of our basic assumptions is that stellar populations in the Solar Cylinder follow a single and relatively narrow AMR. It is interesting to check whether this assumption is compatible with the observed vertical trends of stellar metallicities.

\begin{figure}
    \centering
    \includegraphics[width=\columnwidth]{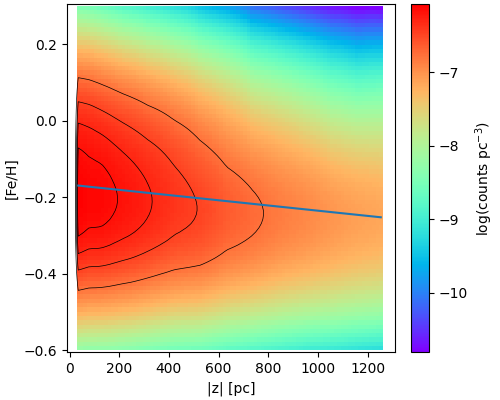}
    \caption{Counts per cubic parsec of giant stars at different $|z|$ and [Fe/H], obtained from the best fit solution, including the spread of 0.1 dex along the metallicity axis. The blue line represents the linear fit to the mean [Feh/] computed in each bin of $|z|$, with a slope of $-0.068\, \text{dex}\,\text{pc}^{-3}$.}
    \label{fig:feh_gradient}
\end{figure}

For instance let us try to reproduce the metallicity gradient found by \citet{imig23} from the last release of the Apache Point Observatory Galactic Evolution Experiment (APOGEE) data. They have considered a high-quality subsample of APOGEE targets defined by $1<\logg[\text{cm}\,\text{s}^{-2}]<2$, $3500<\Teff/\mathrm{K}<5250$. Stars with these characteristics are all red giants, but they do not cover a well-defined range in \gaia magnitudes and colours. Therefore, to derive the metallicity distribution for these stars, we build a new set of partial models comprising only the stars in these $\logg$ and $\Teff$ intervals, for exactly the same population parameters as in our previous analysis (see Sect.~\ref{sec:pmsin}). We then use our best-fitting solution to derive numbers of expected APOGEE targets at every $|z|$ and \feh\ bin, hence constructing the 2D histogram shown in Fig.~\ref{fig:feh_gradient}. The blue line in the Figure shows the linear fit to the mean \feh\ in each vertical slice:
\begin{equation}
    \feh(z) = -0.1671 -0.068 \times |z|
\end{equation}
and therefore the vertical metallicity gradient we obtain is $-0.068\, \text{dex}\,\text{kpc}^{-1}$. Additionally, for the entire Solar Cylinder we find that the mean metallicity is $\feh=-0.19$.

Although we infer a vertical metallicity gradient significantly smaller than the $-0.315\pm0.009\,\text{dex}\,\text{kpc}^{-1}$ mean value found by \citet{imig23} near the Solar Neighbourhood, this exercise demonstrates that a vertical metallicity gradient naturally comes out from the fitting of the SFH in the Solar Cylinder, even when a single AMR is assumed, as the result of the systematic increase in scale height of older (and metal poorer) populations.

Also, it is important to note that APOGEE does not sample all stars in the $\logg$ and $\Teff$ intervals selected for this test. Selection of APOGEE primary targets \citep{zasowski13} is based mainly on 2MASS photometry and it is limited by the apparent brightness of the available targets and by the finite number of observing nights and fibers in the APOGEE spectrograph. Moreover, APOGEE can clearly distinguish populations according to their $[\alpha/\text{Fe}]$, which is a capability still not implemented in our CMD fitting method. Therefore, our attempt to reproduce their metallicity gradient can only be approximate at the moment. Work is underway to fully implement APOGEE target selection in our method, in order not only to verify the abundance gradients derived by the method, but also to actually use their metallicity information during the CMD fitting.

It is also worth mentioning that slightly smaller values for the vertical metallicity gradient have been found with different surveys, for instance \citet{onaltas16} find $-0.157\pm0.003\,\text{dex}\,\text{kpc}^{-1}$ from red clump stars within $|z|<3$~kpc in RAVE, \citet{duong18} find $-0.18\pm0.01\,\text{dex}\,\text{kpc}^{-1}$ from thin disk stars (low-$\alpha$) in GALAH, \citet{hawkins23} finds $-0.15\pm0.01\,\text{dex}\,\text{kpc}^{-1}$ using OBAF stars from LAMOST.

\section{Spatially-correlated SFHs}
\label{sec:correlation}

The present results, illustrated in Fig.~\ref{fig:SFR_2D}, tell of a SFH that, despite being characterized by some noise, is strongly correlated among different locations in the Solar Cylinder, especially at old ages. The spatial correlation occurs at lengths that are comparable to the scale heights we have discussed in Sect.~\ref{sec:heightXage}. It has also a very simple interpretation: \textit{the stellar populations we see in different slices are partially the same, and even more so at old ages}.

Can this simple observation be used to improve the derivation of the SFHs, perhaps allowing us to reduce the noise that characterises the SFH maps of Fig.~\ref{fig:SFR_2D}? To answer this question, we developed the alternative formalism that is described here.

\subsection{Adopting a spatial correlation: the method}
To include a term for the spatial correlation of the solutions at different $z$, we do not need to heavily change our method.
In particular, we can add a term to the likelihood that accounts for the correlation of the SFRs, favouring more correlated solutions.
We take inspiration from Gaussian Processes, and we build a correlation matrix $\bm{\mathsf{R}}$ whose entries are defined by an exponential kernel
\begin{equation}
    \bm{\mathsf{R}}_{n,m;i} = r(\bm{\mathit{x}}_n, \bm{\mathit{x}}_m, l_{i}) = \exp \left\lbrace -\frac{|\bm{\mathit{x}}_n-\bm{\mathit{x}}_m|^{2}}{2l_{i}^{2}} \right\rbrace
\end{equation}
where $\bm{\mathit{x}}_n$ and $\bm{\mathit{x}}_m$ are vectors of coordinates identifying the centre of two slices, and $l_{i}$ is the correlation length. 

The subscript $i$ marks the dependence on age of the correlation length, in fact we assume that each age bin has a different value and that it should be increasing with age. In other words, we are assuming that, from their birth, stellar populations have travelled and got gradually mixed (or spatially distributed) over a distance of the order of the correlation length.
Importantly, it should be noted that we do not attempt to derive the correlation lengths $l_{i}$, but we adopt the results we get from the exponential fit of the SFR presented above ($h_{z,\exp}$ in Table~\ref{tab:result_hz}).
Finally, we compute the covariance matrix for each age bin as
\begin{equation}
    \pmb{\Sigma}_{n,m} = \sigma_{n} \sigma_{m} \bm{\mathsf{R}}
\end{equation}
where $\sigma_{n}$ and $\sigma_{m}$ are the noise standard deviations for the slice $n$ and $m$.

The likelihood in Eq.~\ref{eq:likelihood} is updated to the following:
\begin{multline}
    \log\mathcal{L}(\theta) = \sum_{n=1}^{N_{\mathrm{slice}}} \sum_{j=1}^{N_{\mathrm{Hess}}} \bigl( - M_{n,j}(\theta) + O_{n,j} \log\left(M_{n,j}(\theta)\right) + \\
    - \log(O_{n,j}!) \bigr) + \log\mathcal{L}_{\mathrm{corr}}(\bm{\mathit{a}}_{0},\pmb{\Sigma};\bm{\mathit{a}})
    \label{eq:likelihood_corr}
\end{multline}
where $\bm{\mathit{a}}$ is the vector containing the SFR of all slices and $\log\mathcal{L}_{\mathrm{corr}}(\bm{\mathit{a}}_{0},\pmb{\Sigma};\bm{\mathit{a}})$ is the log-likelihood of the correlated $\bm{\mathit{a}}$, that is the log-likelihood of a multivariate normal distribution centred on $\bm{\mathit{a}}_{0}$ and having covariance matrix $\pmb{\Sigma}$.

\subsection{Adopting a spatial correlation: the results}
\begin{figure}
    \centering
    \includegraphics[width=0.90\columnwidth]{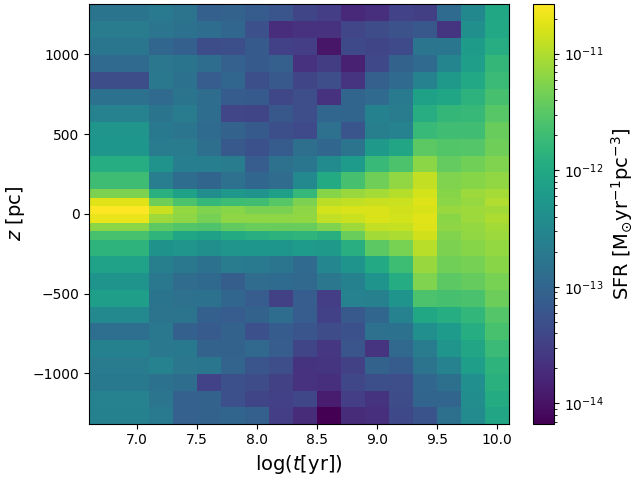}
    \caption{Plot of the SFR per unit volume, as determined for all slices and all age bins in the correlated case.}
    \label{fig:SFR_2D_corr}
\end{figure}
\begin{figure}
    \centering
    \includegraphics[width=0.90\columnwidth]{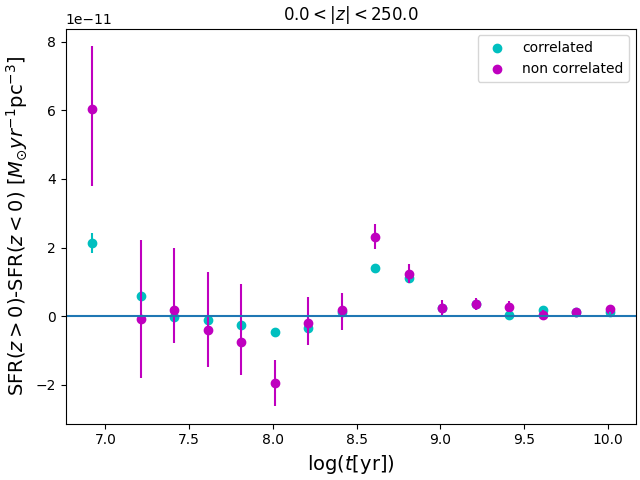}\\
    \includegraphics[width=0.90\columnwidth]{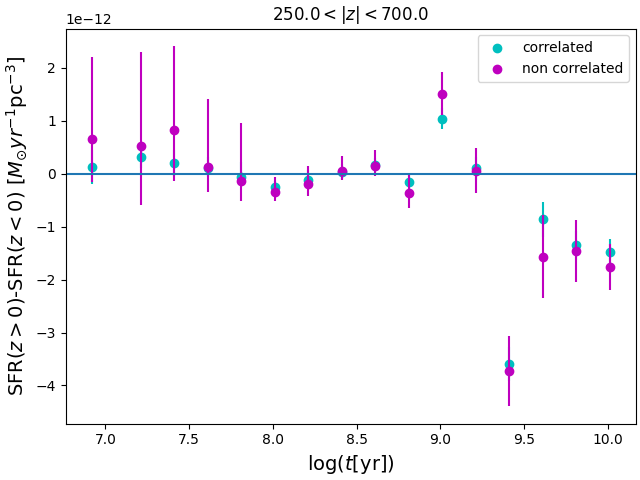}\\
    \includegraphics[width=0.90\columnwidth]{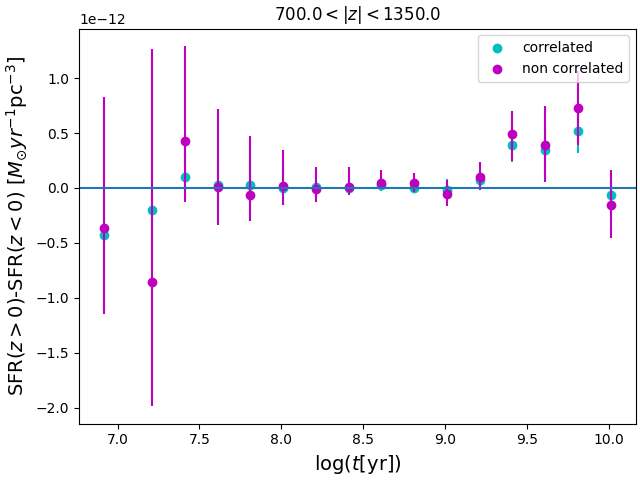}\\
    \caption{Difference between the fitted star formation rate above ($z>0$) and below ($z<0$) the Galactic Plane in the uncorrelated (magenta points) and correlated (cyan points) for three groups of slices: $0.0<z<250.0$ (top panel), $250.0<z<700.0$ (middle panel), $700.0<z<1350.0$ (bottom panel).}
    \label{fig:sfr_corr_nocorr}
\end{figure}
\begin{figure*}
    \centering
    \begin{tabular}{cccc}
        \includegraphics[width=0.90\columnwidth]{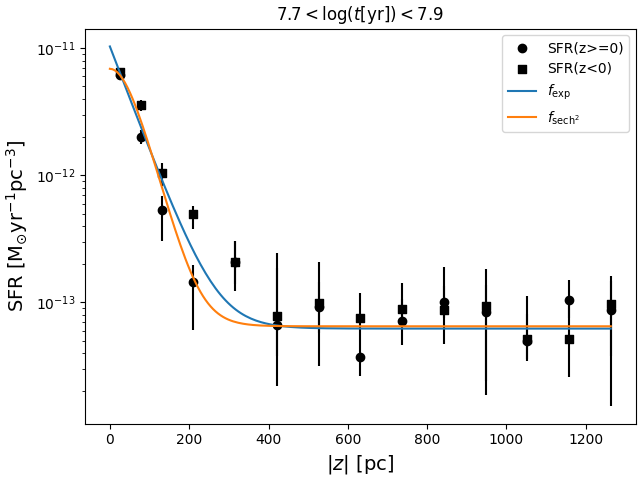}
        & \includegraphics[width=0.90\columnwidth]{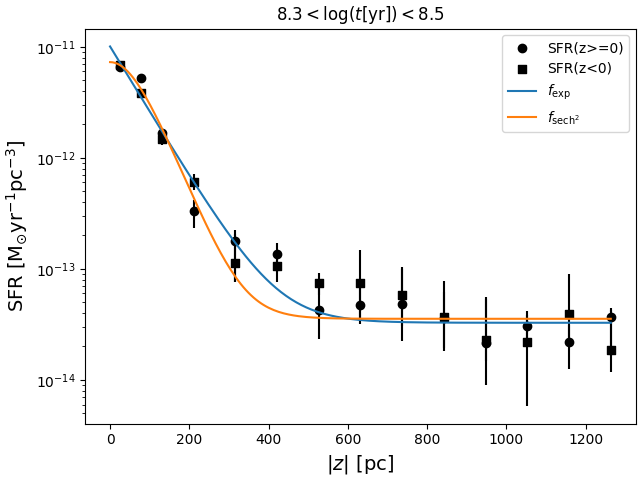}\\
        \includegraphics[width=0.90\columnwidth]{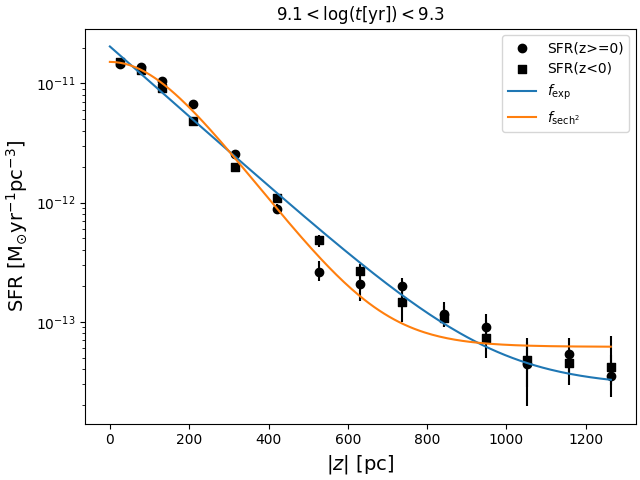}
        & \includegraphics[width=0.90\columnwidth]{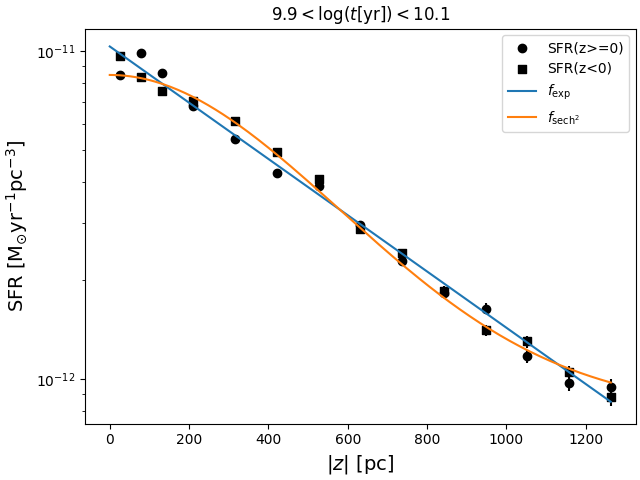}\\
    \end{tabular}
    \caption{Fit of the SFR per unit volume (black circles and squares) with the exponential (blue) and squared hyperbolic secant (orange) functions in the case with correlation.}
    \label{fig:sfr_hz_fit_corr}
    \begin{tabular}{cc}
        \includegraphics[width=0.90\columnwidth]{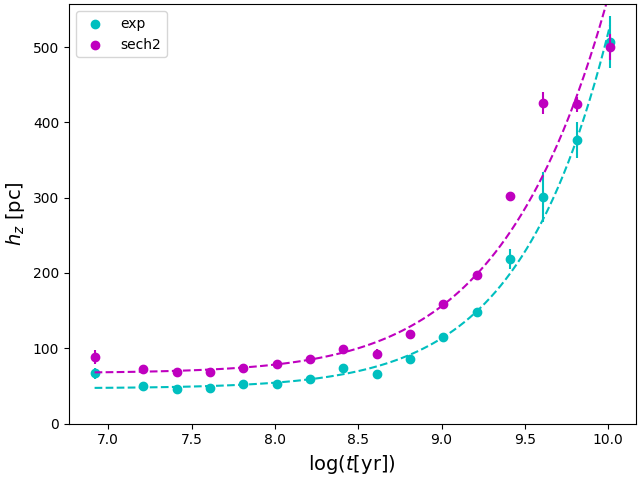}
        & \includegraphics[width=0.90\columnwidth]{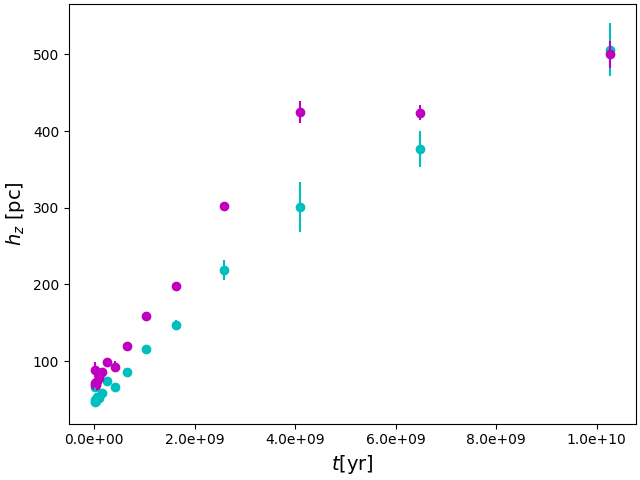} \\
    \end{tabular}
    \caption{The scale height of the Galactic Disc $h_{z}$ versus the mean age of every age bin in the case with correlation. The cyan points are the exponential case, while the magenta points are for the case of the square hyperbolic secant. The left plot uses a logarithmic scale for the age, while the right one uses a linear one and shows that, just as in the case without correlation, the linear trend in younger ages is lost at older ages, where the trend in $h_{z}$ flattens.}
    \label{fig:hz_vs_age_corr}
\end{figure*}
The result of adding the correlation term to our analysis is presented in Fig.~\ref{fig:SFR_2D_corr}.
Comparing this result to Fig.~\ref{fig:SFR_2D} it is possible to appreciate that, while it maintains unchanged the overall trend, it shows a smoother distribution of SFR in the slices closer to the Galactic Plane at all ages, and does not show any more the SFR peak at the youngest age in the slice just above the Galactic Plane.
Furthermore, the distribution of SFR appears to be more symmetric with respect to the Plane. This is presented in Fig.~\ref{fig:sfr_corr_nocorr} which shows the difference between the SFR above and below the Plane in the uncorrelated and correlated cases for three groups of slices: $0.0<|z|<250.0$ (top panel), $250.0<|z|<700.0$ (middle panel), $700.0<|z|<1350.0$ (bottom panel). In all three panels, the values of $\mathrm{SFR}(z>0)-\mathrm{SFR}(z<0)$ in the correlated case appear to be compatible with those of the uncorrelated one and they are also generally closer to zero, thus showing a more symmetric picture of the Disc. In addition, the correlated solution, due to the smaller $1\sigma$ uncertainties, can be considered more reliable than the uncorrelated one.

Figure~\ref{fig:sfr_hz_fit_corr} shows the fits of the SFR at different age bins with eqs.~\ref{eq:sfr_exp} and \ref{eq:sfr_sech2}. The differences with the uncorrelated case are not extreme, but it can be appreciated that the spread of the SFR at each age has decreased.

Looking at the values of $h_{z}$, in Fig.~\ref{fig:hz_vs_age_corr}, the general trend is the same, but the value of $h_{z}$ at the fifth age bin ($7.7 < \logtyr <7.8$) has decreased considerably and is now more in line with the ones found at the young ages.
With regard to Eq.~\ref{eq:hzt}, in this case we obtain
\begin{equation}
    \begin{split}
        h_{z,0} &= 44.8 \pm 2.3~\mathrm{pc}\\
        \tau &= (4.30 \pm 0.99) \times 10^{8}~\mathrm{yr}\\
        n &= 0.766 \pm 0.059
    \end{split}
\end{equation}
for Eq.~\ref{eq:sfr_exp} and
\begin{equation}
    \begin{split}
        h_{z,0} &= 65.9\pm 3.5~\mathrm{pc}\\
        \tau &= (3.4 \pm 1.0) \times 10^{8}~\mathrm{yr}\\
        n &= 0.631 \pm 0.063
    \end{split}
\end{equation}
for Eq.~\ref{eq:sfr_sech2}, and the left panel of Fig.~\ref{fig:hz_vs_age_corr} shows the two fits as dashed lines.
Comparing Fig.~\ref{fig:hz_vs_age} with Fig.~\ref{fig:hz_vs_age_corr}, it can be appreciated that the results in the case including the correlation have smaller fit residuals compared to the non-correlated ones.

As a concluding remark, it should be noted that we have also implicitly adopted a perfect correlation between different slices, in the form of imposing that $\deltacol$, $\deltamag$, $\Delta\feh$ and $\fbin$ values are the same in all slices -- which is already an improvement over current methods -- while we use the flexible correlation on the $a_{i}$ coefficients.
This is due to the assumption that populations seen at different $z$ are the superposition of the same building blocks. A full correlated treatment could be implemented by applying the same kind of correlation to all parameters involved.

Overall, the changes brought by the correlation are not extreme and are compatible with the non-correlated one, but at the same time the overall picture improves.
The approach that we develop and apply for the first time in this work can be extended to other targets, and most importantly to other galaxies, as long as enough photometric data and extensive ASTs are available.

 \section{conclusions}
 \label{sec:conclusions}

In this work, we perform a simultaneous analysis of the SFH across several slices of a cylinder 2.6 kpc high and with a 200~pc radius centred on the Galactic Plane close to the Sun. Our method is based on the fitting of the star counts across the CAMD obtained from \gaia data -- a method routinely used in external galaxies with high-precision photometry. Just a few adaptations were needed in order to face the complexities of \gaia data; in particular we introduce a more careful approach to dealing with binary systems and their resolvability into two components as a function of distance, and do our best to simulate the incompleteness, parallax and photometric errors present in \gaia data. Model parameters that should be common to all slices, such as the possible systematic shifts in colour and magnitude, the initial binary fraction, and the AMR, are all adjusted at the same time. Among our most relevant findings,
 we highlight the following aspects:
\begin{itemize}
    \item We derive the spatial density of SFR as function of both age and height from the Galactic Plane. 
    \item Our results provide an improved picture of the SFH with respect to  previous SFH papers based on \gaia. In particular, we clearly see the spatial dependence of the SFH with the height $z$ above/below the Galactic Plane.
    \item Our space-resolved SFHs also allow us to quantify the increase of scale height with age, in a way that was not possible before: using the entire section of the CAMD which is sensitive to age, instead of a just few selected stellar tracers. A better resolution in age is expected with this method. We confirm that fitting star counts across the SGB section of the CAMD generally produces similar results for the scale height at ages larger than 2 Gyr. However, the same does not happen with the red clump, which just provides an age-averaged value of scale height.
    \item Some details of the results are weird, as for instance the persistent low-level SFR at young ages and very far from the Galactic Plane, which is indicated by a non-negligible value of the constant $c$ in our fits of the scale height. These probably indicate shortcomings of stellar models and/or of \gaia data. Identifying these problems (rather than hiding them) is a necessary step to further improve the description of nearby stellar populations in the post-\gaia era.
    \item The surface density of stellar formation as a function of age, $\Psi(t)$, is derived in three different ways. These functions clearly reveal a period of enhanced star formation in the age bin going from 2 to 3 Gyr. This peak has some correspondence with peaks inferred by other authors at similar ages \citep{mor19,ruiz22,syso22} but using different methods and data. Other peaks detected by these authors, however, do not appear in our results.
    \item The total mass of stars formed per unit area which we determine is roughly double that determined by studies using kinematic information within $1$~kpc from the Galactic Plane. This might be implying a high degree of matter recycling in successive generation of stars.
    \item The parameters derived from our analysis -- comprising $\Psi(t),h_z(t),\mathrm{[Fe/H]}(t),\fbin$ -- can be easily implemented in our population synthesis code TRILEGAL \citep{girardi05}, hence allowing us to perform better simulations of the Galactic Disc for different spatial samples and in other photometric systems. Indeed, a web interface incorporating these results\footnote{\url{http://stev.oapd.inaf.it/trilegal}} is presently in preparation.
    \item Finally, we modify the SFH-recovery method to take advantage of the very evident spatial correlation in the old SFH revealed by our analysis. An age-dependent correlation length is introduced along with its corresponding covariance matrix and changes in the computation of the model--data likelihood. The modified method is shown to provide less noisy and more symmetric SFH results with respect to the Galactic Plane. We consider it as a promising way to improve SFH-recovery methods to be applied to nearby galaxies.
\end{itemize}

This work represents an intermediate step within a more ambitious project to improve the tools for the modelling and the interpretation of stellar populations in disc galaxies. Next steps of this project will include: a recalibration of the parameters defining the populations of binaries, the exploration of a large range in Galactocentic radius, and the inclusion of spectroscopic and asteroseismic information into our modelling.

\section*{Data Availability}
The data underlying this article were accessed from the ESA \gaia archive (\url{https://gea.esac.esa.int/archive/}). The best-fitting models generated in this research will be shared on reasonable request to the corresponding author.

\section*{Acknowledgements}
We thank the anonymous referee for the useful report.
Alessandro Mazzi acknowledges financial support from Padova University, Department of Physics and Astronomy Research Project 2021 (PRD 2021), and thanks the Center for Computation Astrophysics for the kind hospitality in New York.
Léo Girardi acknowledges partial funding by an INAF Theory Grant 2022.
Andrea Miglio acknowledges support from the ERC Consolidator Grant funding scheme (project ASTEROCHRONOMETRY, G.A. n. 772293 \url{http://www.asterochronometry.eu}).
The Flatiron Institute is funded by the Simons Foundation.

We are grateful to the entire \texttt{GaiaUnlimited} team for their software and for organising this very helpful workshop.

\bibliographystyle{mnras}
\bibliography{gaia_cmd}

\clearpage
\appendix

\section{Example of a query}
\label{app:query}
We show below a template of the query we performed on the Gaia archive to select the quantities of interest for all sources within a distance $\rmax$ from the Sun:

\noindent\footnotesize{\texttt{select ra, dec, l, b, ecl\_lat, astrometric\_params\_solved, parallax, parallax\_error, parallax\_over\_error, astrometric\_excess\_noise, astrometric\_excess\_noise\_sig, ruwe, phot\_g\_mean\_mag, phot\_g\_mean\_flux, phot\_g\_mean\_flux\_error, phot\_bp\_mean\_mag, phot\_bp\_mean\_flux, phot\_bp\_mean\_flux\_error, phot\_rp\_mean\_mag, phot\_rp\_mean\_flux, phot\_rp\_mean\_flux\_error, phot\_bp\_rp\_excess\_factor, bp\_rp, pseudocolour, nu\_eff\_used\_in\_astrometry from gaiadr3.gaia\_source where 1000/parallax<$\rmax$}}

\section{Derivation of errors for ASTs}
\label{app:errors}
\begin{figure*}
    \centering
    \begin{adjustbox}{center}
        \setlength\tabcolsep{1.5pt}
        \begin{tabular}{cc}
            \includegraphics[width=0.85\columnwidth]{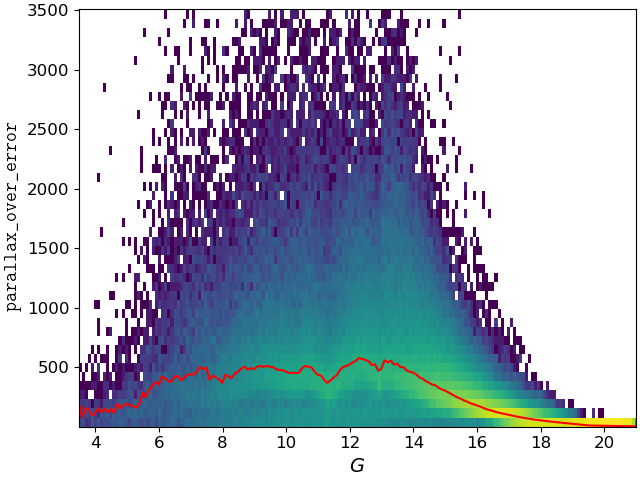} &
            \includegraphics[width=0.85\columnwidth]{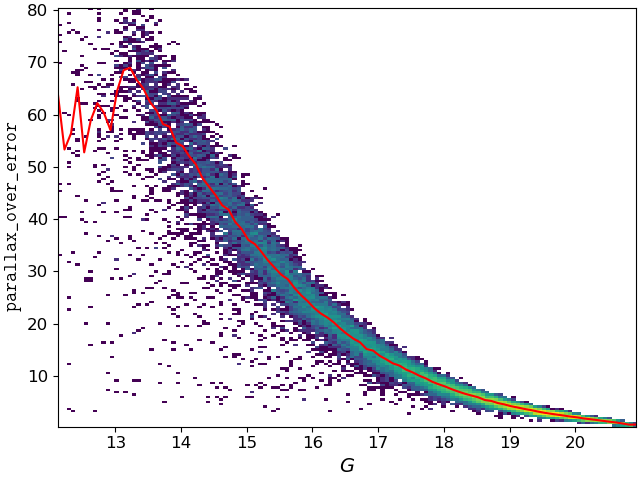} \\
            \includegraphics[width=0.85\columnwidth]{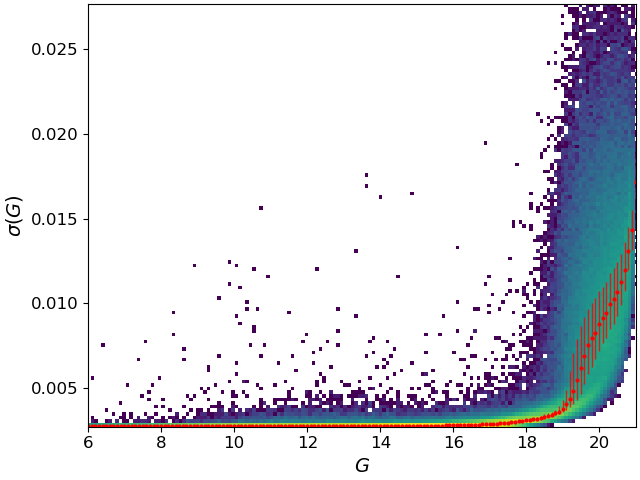} &
            \includegraphics[width=0.85\columnwidth]{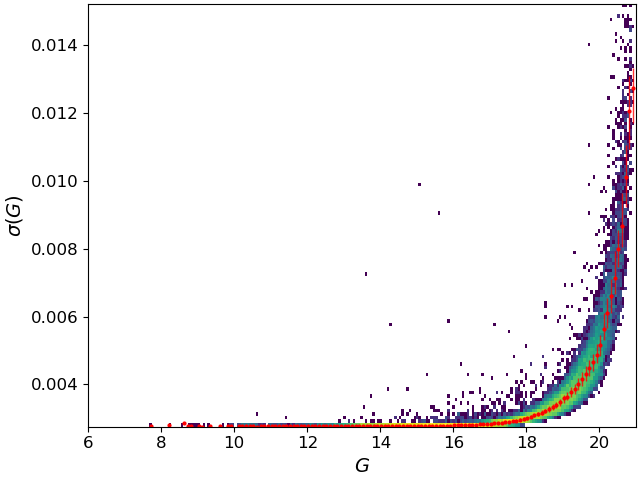} \\
            \includegraphics[width=0.85\columnwidth]{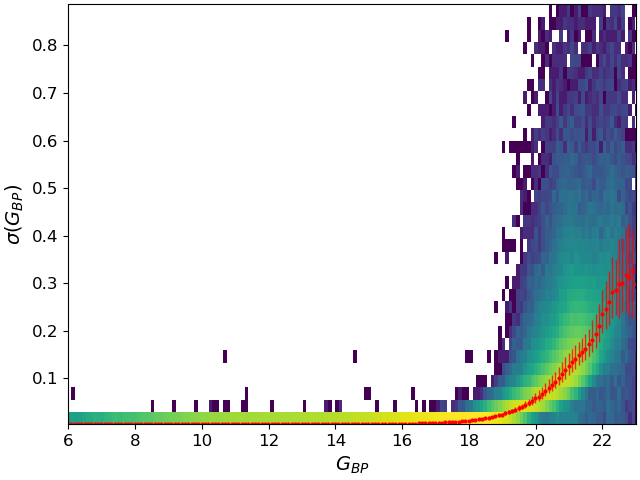} &
            \includegraphics[width=0.85\columnwidth]{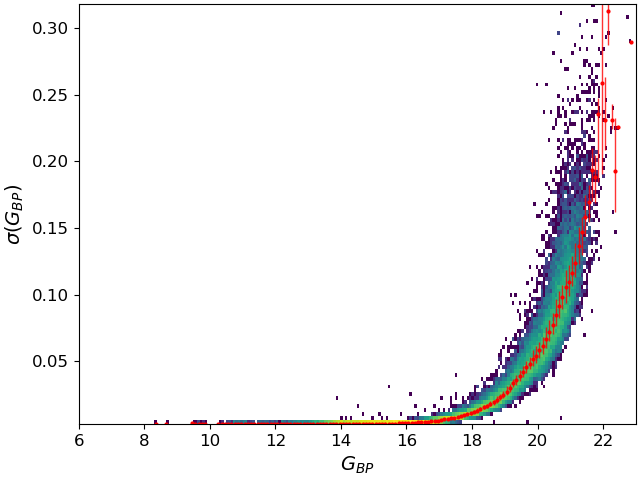} \\
            \includegraphics[width=0.85\columnwidth]{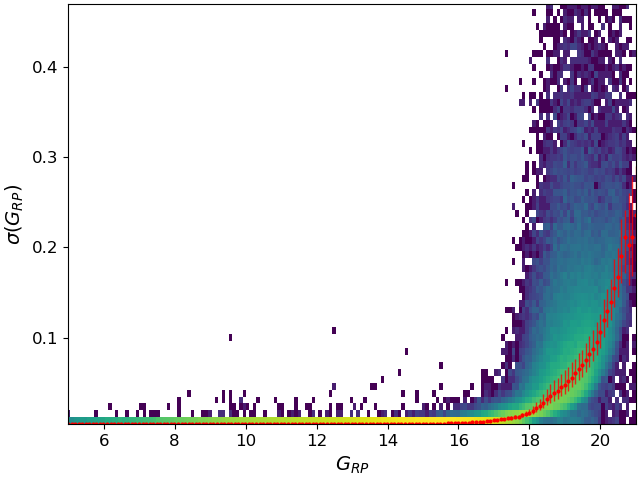} &
            \includegraphics[width=0.85\columnwidth]{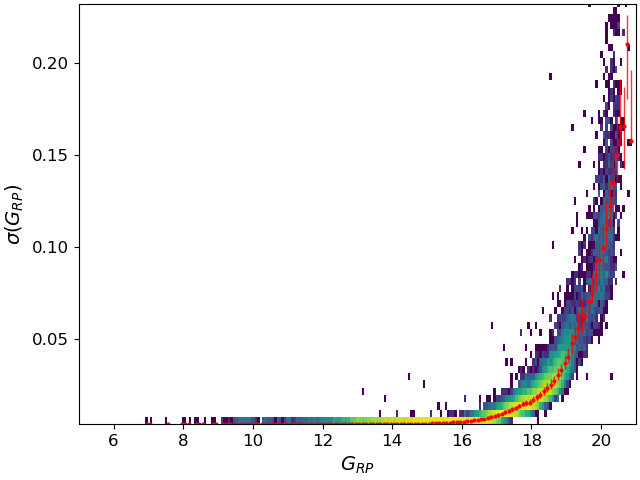} \\
        \end{tabular}
    \end{adjustbox}
    \caption{Distribution of errors (2D histograms) and their median values in bins of $0.1$~mag (red lines) for the $0.00~\text{pc} < z < 52.63~\text{pc}$ slice (left column) and $894.74~\text{pc} < z < 1000.00~\text{pc}$ one (right column). \textit{First row}. \gaia's \texttt{parallax\_over\_error}. \textit{Second, third and fourth rows, respectively}. $\gmag$, $\gbp$ and $\grp$ magnitude errors.}
    \label{fig:errors}
\end{figure*}
Here we detail how we compute the errors needed for our AST-like processing described in Section~\ref{sec:models}.

To determine the astrometric errors to apply to a fake star with parallax $\pi$ and apparent magnitude $\gmag$ belonging to a slice $n$, we start by binning the distribution of \gaia DR3 \texttt{parallax\_over\_error} in the slice with $0.1$~mag wide \gmag\ bins. Then, we compute the median value in each bin. The results are illustrated in the leftmost panels of Fig.~\ref{fig:errors} for the slices $0~\text{pc}<z<52.63~\text{pc}$ and $894.74~\text{pc}<z<1000~\text{pc}$.

The same procedure is followed for the photometric errors in the $\gmag$, $\gbp$ and $\grp$ magnitudes. However, the magnitude errors are not provided in the \gaia DR3 catalogue; they are instead derived from the flux errors \texttt{phot\_X\_mean\_flux}, where \texttt{X} is one of \texttt{g}, \texttt{bp} and \texttt{rp}. In particular, we use
\begin{equation}
    \sigma(\texttt{X}) = \sqrt{ \left( \frac{-2.5}{\log(10)} \frac{\sigma(F_\texttt{X})}{F_\texttt{X}} \right) ^{2} + Z_\texttt{X}^{2}}
\end{equation}
where $F_\texttt{X}$ is the flux, $\sigma(F_\texttt{X})$ is its corresponding error, and $Z_\texttt{X}$ is the zeropoint for each band.

To determine the errors of an artificial star with randomly generated parallax and magnitudes, we proceed as follows.
\begin{itemize}
    \item For the astrometric errors, we first interpolate between the median values of \texttt{parallax\_over\_error} at the position of the artificial star's $\gmag$ magnitude. Then, we use this value as the standard deviation of a Gaussian centred on zero to determine the artificial star's value of \texttt{parallax\_over\_error}, and we multiply its parallax by the inverse of this quantity.
    \item For the photometric errors, for each band we interpolate between the median values of $\sigma(\texttt{X})$ using the value of the star's fake \texttt{X} magnitude.
\end{itemize}

\section{Example of single and binary partial models.}
\label{app:pms}
\begin{figure*}
    \centering
    \includegraphics[height=0.5\textwidth, trim={0  30 65 65}, clip]{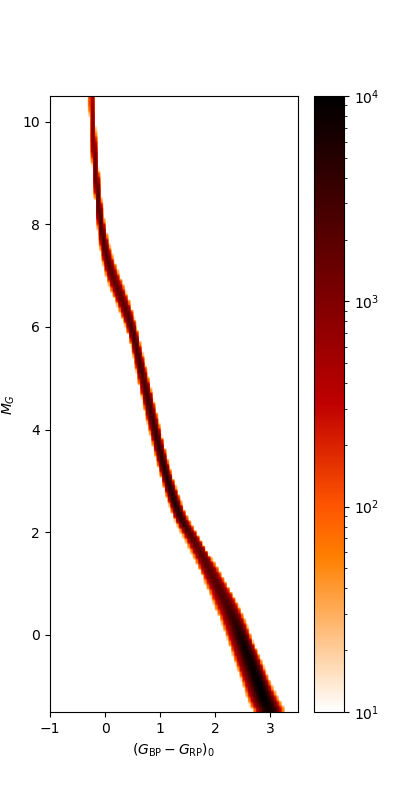}
    \includegraphics[height=0.5\textwidth, trim={35 30 65 65}, clip]{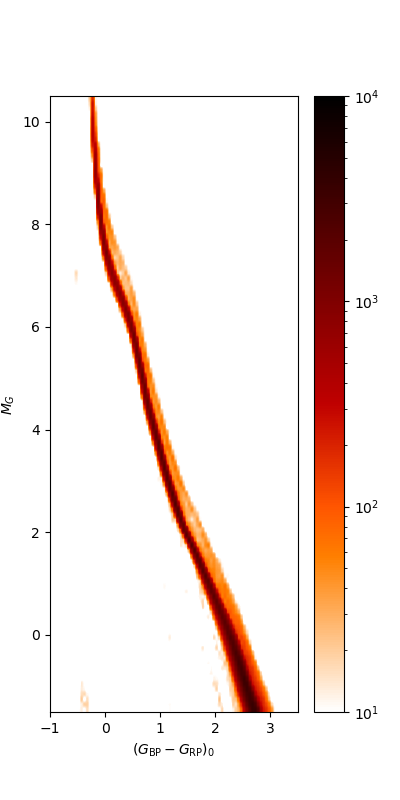}
    \includegraphics[height=0.5\textwidth, trim={35 30 65 65}, clip]{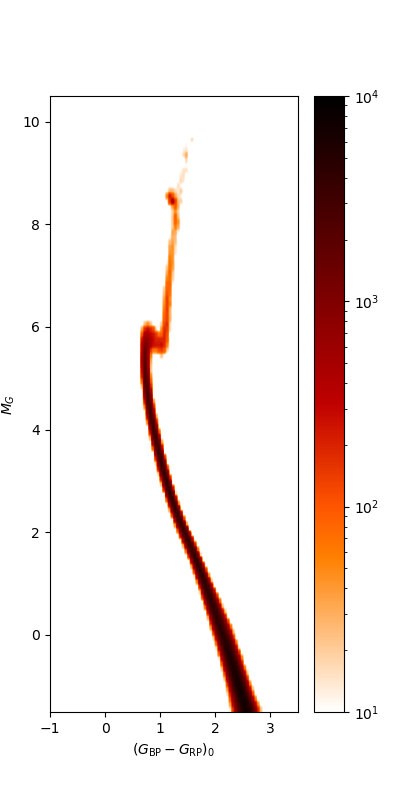}
    \includegraphics[height=0.5\textwidth, trim={35 30 65 65}, clip]{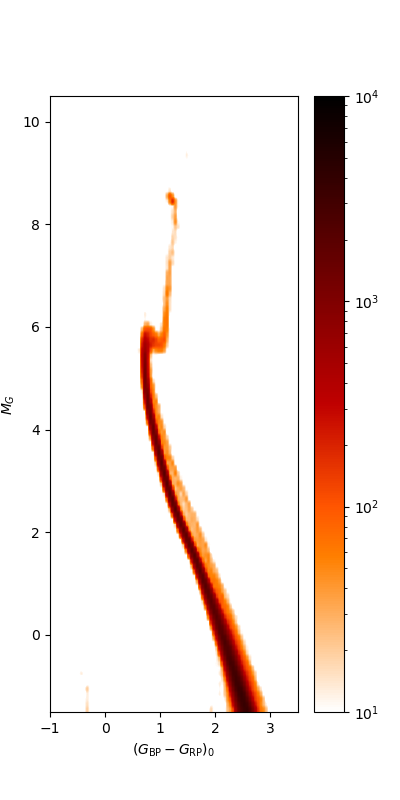}
    \caption{Examples of partial models for young ($7.70<\logtyr<7.90$; left panels) and old ($9.70<\logtyr<9.90$; right panels) age bins, and for single (first and third panels) and binary (second and fourth panels) stars. They refer to the models along the reference AMR and for the slice closest to the Sun. For every age bin, the total initial stellar mass represented by the single and binary partial models is the same. These particular models are computed for the reference AMR (that is, they are in the PM$_i^0$ sequence), and include the modest photometric errors and incompleteness of the slice closest to the Sun.}
    \label{fig:example_pms}
\end{figure*}
In Fig.~\ref{fig:example_pms} we show examples of single and binary partial models in two age bins, namely $7.70<\logtyr<7.90$ and $9.70<\logtyr<9.90$, taking into account the incompleteness and photometric errors of the slice that comprises the Sun.
It can be noticed that the main differences between single and binary partial models are in the broader main sequence and the smaller numbers of objects present in the binary models. Indeed, single and binary partial models in this figure correspond to the same total initial mass in stars. Before the SFH determination is started, these Hess diagrams are re-scaled so that they will represent a constant SFR of 1~$\Msun\,\mathrm{yr}^{-1}$.

\section{On the metallicity interval covered by our AMR}
\label{app:AMR}
\begin{figure*}
    \centering
    \includegraphics[width=0.65\textwidth, trim={30 4 55 20}, clip]{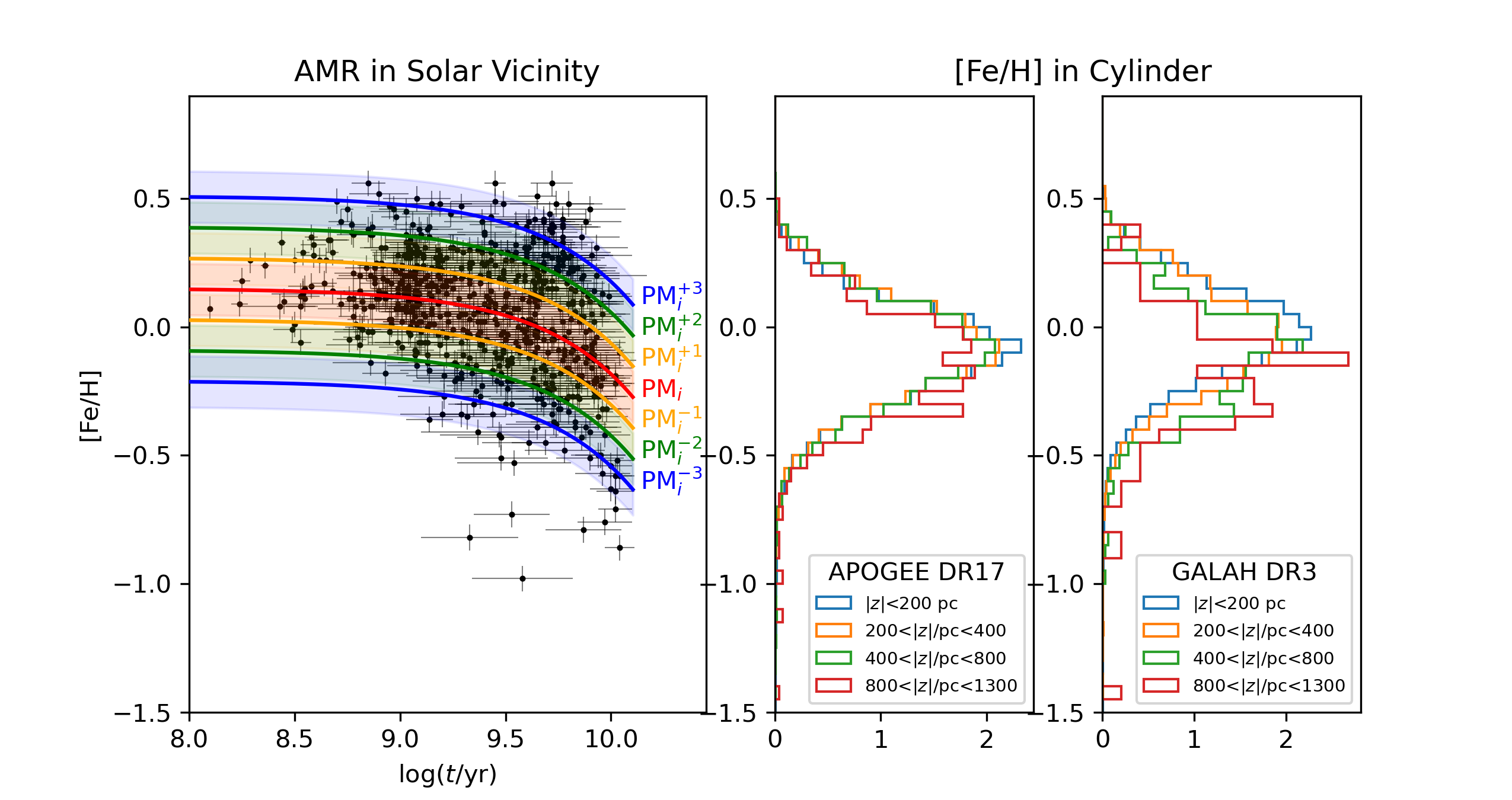}
    \caption{\textit{Left panel}. Our reference AMR (red line) and the AMR of the additional partial models we use in this work. Every partial model spans a distribution of metallicities defined by a Gaussian whose $1\sigma$ values are illustrated by the colour-shaded areas. These partial models are compared to the high-quality sample of stars from \citep{feuillet16}, with metallicities from APOGEE \citep{apogee, aspcap} and ages estimated from their position on the HR diagram. 97 per cent of these stars are within 300 pc from the Sun. One can notice that the overall interval of age-metallicity plan that our code is allowed to explore contains the vast majority of these stars. \textit{Middle and right panels}. Histograms of the metallicities measured in stars contained in our Cylinder by the APOGEE \citep[DR17; ][]{apogee_dr17} and GALAH \citep[DR3;][]{galah_dr3} surveys, respectively. The data in these latter cases comes from the \texttt{StarHorse} catalog from \citet{queiroz23}, with distances based on \gaia DR3. The data are separated in four bins of $|z|$ and the histograms are re-normalized to 1, highlighting the weak dependence of the metallicity distribution on $|z|$.}
    \label{fig:RAMR}
\end{figure*}

The reference AMR and the partial models adopted in this work are the same as in the analysis of a sample within 200 pc by \citet{daltio21}. Having we significantly expanded the volume covered by the data, it is  necessary to verify if the range of metallicities we adopt is still suitable for the SFH analysis.

The left panel of Fig.~\ref{fig:RAMR} compares our partial models in the age--metallicity plane with the AMR inferred from APOGEE data for the Solar Neighbourhood by \citet{feuillet16}. This sample is concentrated at distances smaller than 300~pc, and it is evident that our partial models are describing the bulk of such data.

As we increase the distance from the Galactic midplane, age--metallicity relations become less reliable, but we can at least verify the overall changes in metallicity distributions. This is illustrated in the middle and right panels of Fig.~\ref{fig:RAMR}, which present histograms of the metallicities measured for stars in our cylinder by two high-resolution spectroscopic surveys, namely APOGEE DR17 \citep{apogee_dr17} and GALAH DR3 \citep{galah_dr3}. These surveys adopt quite different wavelength ranges and strategies for target selection and data analysis. As can be noticed, in both cases the samples farther to the Sun tend to be metal poorer. However, the mode of the distributions change by just $\sim0.2$ dex across our 1300~pc interval of $|z|$. The total metallicity interval also changes with $|z|$, but the bulk of stars are always located inside the $-0.6<\feh<+0.4$ interval. In the section of the Cylinder with $800~\mathrm{pc}<|z|<1300~\mathrm{pc}$, the fraction of stars with $\feh<-0.7$~dex amounts to just 1.8 and 3 per cent, in the APOGEE and GALAH surveys, respectively. Stars in this low-metallicity tail of the distribution are the only ones not being described by the partial models used for the Galactic Disc (Sects.~\ref{sec:pmsin} and \ref{sec:pmbin}); however, they can still be described by our partial model for the halo (Sect.~\ref{sec:halo}).

\section{The binary fraction}
\label{app:binary_uncert}
\begin{figure}
    \centering
    \includegraphics[width=0.9\columnwidth]{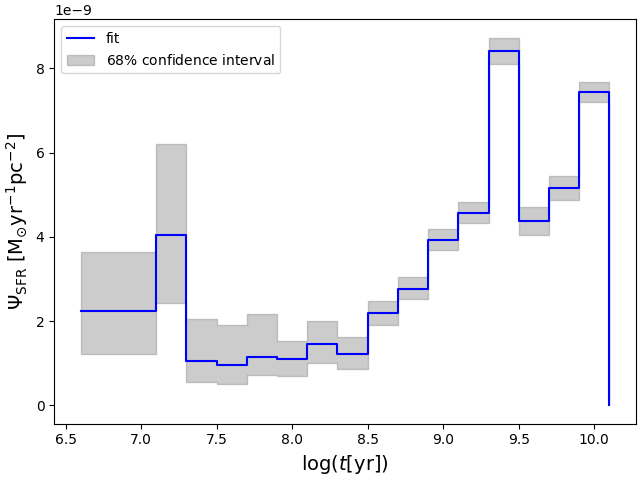}
    \caption{SFR per unit area in the case with the binary fraction fixed to 0.4. Compared to Fig.\ref{fig:sfr_tot}, the shape is almost the same, but the values at each age decrease by 40-60 per cent.}
    \label{fig:sfr_fbin04}
\end{figure}
\citet{daltio21} analysed different parts of the \gaia DR2 CAMD, limited to stars within 200~pc from the Sun and at $|b|>25^\circ$, using two different distributions of initial binary parameters, namely those from \citet{Eggleton2006} and \citet{moe17}. They reach the interesting conclusion that smaller binary fractions, of order of $\fbin\simeq0.4$, are favoured when only the lower main sequence is analysed via the CAMD-fitting method. This conclusion depends on the fraction of twins (i.e. binaries in which both components have a similar initial mass) present in these distributions: these are the binaries that appear as a ``second main sequence'', about 0.7~mag brighter than the single-star main sequence in the \gaia CAMD (see Fig.~\ref{fig:example_pms}). The frequency of twins in itself is a quantity being reevaluated as a consequence of \gaia data \citep[e.g.][]{elbadry19, elbadry21}; therefore the conclusion is that $\fbin\simeq0.4$ should be taken with the due care.

For completeness, in this work we also explore a case with a fixed binary fraction $\fbin=0.4$, while keeping all other aspects of or analysis the same. The fitted SFH in this case behaves in a very similar way to the solutions already presented in this paper. It results in $M_{\mathrm{tot}} = (9.36 \pm 0.45) \times 10^{6} \, \Msun$ and $\Sigma_{\text{SFR}} = 74.47 \pm 0.36\, \Msun \, \mathrm{pc}^{-2}$. The SFR per unit area in all age bins is presented in Fig.~\ref{fig:sfr_fbin04}. Comparing with Fig.~\ref{fig:sfr_tot}, it can be appreciated that the shape is remarkably similar, with just a slight relative decrease at the young ages. The main change in this case is in the absolute scale of the resulting SFR: all SFR values become about 40-60 per cent smaller than in the case with \fbin\ close to 1. This is the consequence of the larger total mass one needs, in a population of binaries (including the unresolved ones), to produce the same number of distinct stellar objects as in a population of single stars.

These aspects will be further investigated in a subsequent work, dealing with the different choices for modelling the IMF and binaries in a high-quality subsample of \gaia data. For the moment, we consider the binary fraction as a possible, additional uncertainty affecting the SFR values we derive from the \gaia CAMD.

\section{Investigating the young SFR}
\label{sec:young_sfr}

As shown by Figs.~\ref{fig:results_comparison} and \ref{fig:result_sfr}, an excess of SFR associated to the youngest age bins can be seen up to the farthest slice from the Galactic Plane. We explored possible solutions to this issue, namely:
\begin{itemize}
    \item using a half-Gaussian prior centred on zero for the SFR instead of a uniform one;
    \item adding the term
    $\mathcal{L}_{\mathrm{extra}} = -\sum_{n}^{N_{\mathrm{slice}}}\sum_{j}^{N_{\mathrm{Hess}}}O_{n,j}$, with $(n,j)$ such that $M_{n,j}=0$, to the likelihood in Eq.~\ref{eq:likelihood}, so as to take into account the cells where M=0 and O>0, which are not considered otherwise;
    \item modifying M and O in Eq.~\ref{eq:likelihood} by adding a background of star counts of $10^{-10}$, $10^{-4}$ and $10^{-1}$, such that all cells can contribute to the likelihood.
\end{itemize}

In all the cases discussed above, the result presented always the excess of young SFR at high values of $|z|$.
Based on these experiments, we conclude that this excess is really required by the method. Young populations are added for two reasons. First, because there are, in most slices, some bright main sequence stars that cannot be fitted with purely-old populations (see Table \ref{tab:counts_zlice1}). Second, even in the cases when there are no bright main sequence stars, the addition of young populations may be increasing the likelihood computed for lower parts of the CAMD more than they decrease the likelihood computed at the top part of the main sequence. It would be possible to get rid of these undesired young contributions by choosing a convenient prior for the final SFR, but we prefer not to do so because, in the end, the problem is minor, and it is probably signalling small inadequacies either in the \gaia data, or in the stellar models being adopted to model it.

\section{Additional tables}
In this section we present some tables that might be useful to the general reader, and which are intended to become online-only material.

\begin{table*}
    \caption{Counts of stars in the CAMD within small age ranges in different slices of our data. The age range is indicated in the header of the columns from the second to the last one, while the boundaries in $z$ of each slice are reported in the first two columns.}
    \label{tab:counts_zlice1}
    \centering
    \setlength\tabcolsep{3.0pt}
    \begin{tabular}{@{\extracolsep{4pt}}cccccccccccc@{}}
        \toprule
        &  & \multicolumn{8}{c}{log($t$[yr]) interval of SGB} \\
        \cline{3-10}
        $\zmin~[\text{pc}]$ & $\zmax~[\text{pc}]$ & 9.3-9.4 & 9.4-9.5 & 9.5-9.6 & 9.6-9.7 & 9.7-9.8 & 9.8-9.9 & 9.9-10.0 & 10.0-10.1 & Young Main Sequence & Red clump \\
        \cline{1-2}
        \cline{3-10}
        \cline{11-11}
        \cline{12-12}
        -1315.78 & -1210.52 & 0 &   9 &  15 &  15 &  19 &  28 &  40 &   5 &    0 &  27 \\
        -1210.52 & -1105.26 & 1 &  12 &  15 &  10 &  21 &  23 &  40 &   9 &    1 &  34 \\
        -1105.26 & -1000.00 & 2 &   4 &  17 &  10 &  22 &  40 &  71 &  18 &    5 &  58 \\
        -1000.00 & -894.74 &  3 &   6 &  18 &  16 &  33 &  41 &  60 &  17 &    9 &  62 \\
         -894.74 & -789.47 &  3 &   9 &  39 &  33 &  41 &  67 &  79 &  21 &    5 &  75 \\
         -789.47 & -684.21 &  4 &  12 &  38 &  40 &  62 &  83 & 105 &  41 &    5 & 110 \\
         -684.21 & -578.95 &  5 &  19 &  81 &  61 &  74 &  97 & 107 &  56 &   10 & 158 \\
         -578.95 & -473.68 & 11 &  45 & 135 & 111 & 143 & 125 & 177 &  77 &   30 & 253 \\
         -473.68 & -368.42 & 24 &  67 & 180 & 158 & 193 & 189 & 223 &  98 &   62 & 363 \\
         -368.42 & -263.16 & 30 &  96 & 275 & 190 & 229 & 205 & 229 & 114 &  118 & 531 \\
         -263.16 & -157.89 & 42 & 107 & 403 & 273 & 279 & 275 & 285 & 116 &  345 & 727 \\
         -157.89 & -105.26 & 25 &  92 & 194 & 143 & 168 & 170 & 148 &  60 &  378 & 478 \\
         -105.26 &  -52.63 & 27 &  82 & 237 & 180 & 160 & 147 & 185 &  76 &  746 & 593 \\
          -52.63 &    0.00 & 46 & 109 & 253 & 187 & 191 & 173 & 174 &  99 & 1102 & 593 \\
            0.00 &   52.63 & 43 & 101 & 283 & 192 & 169 & 174 & 182 &  88 & 1109 & 591 \\
           52.63 &  105.26 & 54 & 111 & 242 & 202 & 158 & 179 & 176 & 101 &  952 & 560 \\
          105.26 &  157.89 & 32 &  90 & 213 & 154 & 187 & 168 & 179 &  77 &  576 & 498 \\
          157.89 &  263.16 & 49 & 141 & 378 & 277 & 296 & 288 & 259 & 124 &  489 & 786 \\
          263.16 &  368.42 & 25 &  85 & 225 & 185 & 207 & 216 & 226 &  58 &  181 & 457 \\
          368.42 &  473.68 & 16 &  61 & 169 & 146 & 147 & 137 & 200 &  62 &   54 & 308 \\
          473.68 &  578.95 &  5 &  33 & 109 & 102 & 117 & 127 & 165 &  64 &   23 & 238 \\
          578.95 &  684.21 &  9 &  21 &  92 &  71 &  97 & 124 & 136 &  39 &   14 & 166 \\
          684.21 &  789.47 &  3 &  21 &  43 &  51 &  64 &  78 & 123 &  27 &   11 & 103 \\
          789.47 &  894.74 &  2 &  10 &  37 &  27 &  57 &  53 &  92 &  19 &    3 &  96 \\
          894.74 & 1000.00 &  4 &   3 &  21 &  30 &  49 &  46 &  99 &  12 &    2 &  60 \\
         1000.00 & 1105.26 &  2 &   8 &  15 &  15 &  28 &  34 &  70 &   7 &    4 &  37 \\
         1105.26 & 1210.52 &  2 &   3 &  12 &   9 &  21 &  33 &  33 &   4 &    4 &  28 \\
         1210.52 & 1315.78 &  3 &   7 &   9 &  11 &  11 &  28 &  32 &   7 &    0 &  40 \\
        \bottomrule
    \end{tabular}
\end{table*}

\begin{table*}
\caption{Sample table summarising the results of the SFH analysis in the non-correlated case (Sect.~\ref{sec:results}). A complete table is available in the online material, together with a similar table for the correlated case (Sect.~\ref{sec:correlation}).}
\label{tab:sfr_results}
\begin{tabular}{cc|ccc|ccc|c|ccc}
\hline\hline
$\zmin$ & $\zmax$ & 
$a_1$ & $\sigma(a_1^-)$ &  $\sigma(a_1^+)$ &
$a_2$ & $\sigma(a_2^-)$ &  $\sigma(a_2^+)$ &
$\cdots$ &
$a_{16}$ & $\sigma(a_{16}^-)$ &  $\sigma(a_{16}^+)$ 
\\
\hline
-1315.78 & -1210.52 & 2.492e-04 & 5.884e-05 & 4.558e-05 & 1.965e-04 & 2.514e-05 & 4.999e-05 & $\cdots$ & 8.784e-04 & 4.882e-05 & 5.291e-05 \\
-1210.52 & -1105.26 & 2.625e-04 & 8.987e-05 & 1.098e-04 & 1.706e-04 & 5.211e-05 & 7.440e-05 & $\cdots$ & 1.048e-03 & 4.452e-05 & 4.884e-05 \\
-1105.26 & -1000.00 & 1.920e-04 & 1.116e-04 & 1.180e-04 & 1.543e-04 & 7.948e-05 & 9.524e-05 & $\cdots$ & 1.302e-03 & 5.724e-05 & 5.386e-05 \\
-1000.00 & -894.74 & 2.025e-04 & 1.680e-04 & 7.486e-05 & 2.605e-04 & 1.252e-04 & 1.033e-04 & $\cdots$ & 1.411e-03 & 5.902e-05 & 4.953e-05 \\
-894.74 & -789.47 & 2.487e-04 & 8.466e-05 & 8.090e-05 & 1.865e-04 & 8.692e-05 & 9.689e-05 & $\cdots$ & 1.856e-03 & 6.132e-05 & 6.366e-05 \\
-789.47 & -684.21 & 1.316e-04 & 5.020e-05 & 1.032e-04 & 1.832e-04 & 4.399e-05 & 6.174e-05 & $\cdots$ & 2.418e-03 & 6.866e-05 & 7.319e-05 \\
-684.21 & -578.95 & 3.339e-04 & 1.143e-04 & 9.013e-05 & 1.580e-04 & 4.871e-05 & 8.878e-05 & $\cdots$ & 2.873e-03 & 7.020e-05 & 8.115e-05 \\
-578.95 & -473.68 & 6.773e-04 & 1.327e-04 & 1.253e-04 & 1.621e-04 & 3.283e-05 & 9.552e-05 & $\cdots$ & 4.074e-03 & 7.951e-05 & 8.984e-05 \\
-473.68 & -368.42 & 4.970e-04 & 1.169e-04 & 1.235e-04 & 1.835e-04 & 1.052e-04 & 7.217e-05 & $\cdots$ & 4.926e-03 & 9.424e-05 & 9.195e-05 \\
-368.42 & -263.16 & 7.937e-04 & 1.174e-04 & 1.188e-04 & 2.291e-04 & 6.644e-05 & 6.485e-05 & $\cdots$ & 6.104e-03 & 1.082e-04 & 1.055e-04 \\
-263.16 & -157.89 & 1.431e-03 & 1.695e-04 & 1.545e-04 & 4.638e-04 & 9.982e-05 & 1.268e-04 & $\cdots$ & 7.044e-03 & 1.193e-04 & 1.201e-04 \\
-157.89 & -105.26 & 2.418e-03 & 2.683e-04 & 2.677e-04 & 1.463e-03 & 2.769e-04 & 2.456e-04 & $\cdots$ & 7.569e-03 & 1.699e-04 & 1.584e-04 \\
-105.26 & -52.63 & 6.353e-03 & 5.295e-04 & 5.497e-04 & 3.398e-03 & 3.520e-04 & 3.487e-04 & $\cdots$ & 8.307e-03 & 1.686e-04 & 1.827e-04 \\
-52.63 & 0.00 & 1.997e-02 & 1.348e-03 & 1.491e-03 & 7.988e-03 & 7.468e-04 & 7.776e-04 & $\cdots$ & 9.635e-03 & 1.962e-04 & 1.934e-04 \\
0.00 & 52.63 & 2.657e-02 & 1.968e-03 & 1.895e-03 & 1.339e-02 & 1.116e-03 & 1.224e-03 & $\cdots$ & 8.438e-03 & 1.803e-04 & 1.790e-04 \\
52.63 & 105.26 & 1.772e-02 & 1.350e-03 & 1.270e-03 & 4.243e-03 & 4.659e-04 & 4.605e-04 & $\cdots$ & 9.872e-03 & 2.094e-04 & 1.972e-04 \\
105.26 & 157.89 & 5.161e-03 & 4.143e-04 & 4.639e-04 & 1.287e-03 & 2.617e-04 & 2.612e-04 & $\cdots$ & 8.580e-03 & 1.880e-04 & 1.757e-04 \\
157.89 & 263.16 & 2.002e-03 & 2.081e-04 & 2.179e-04 & 2.170e-04 & 3.992e-05 & 1.243e-04 & $\cdots$ & 6.807e-03 & 1.197e-04 & 1.118e-04 \\
263.16 & 368.42 & 1.138e-03 & 1.593e-04 & 1.510e-04 & 4.778e-04 & 6.202e-05 & 6.827e-05 & $\cdots$ & 5.403e-03 & 1.092e-04 & 9.630e-05 \\
368.42 & 473.68 & 5.165e-04 & 1.208e-04 & 9.228e-05 & 2.095e-04 & 6.324e-05 & 1.187e-04 & $\cdots$ & 4.262e-03 & 8.493e-05 & 8.469e-05 \\
473.68 & 578.95 & 5.056e-04 & 7.996e-05 & 1.202e-04 & 1.937e-04 & 7.660e-05 & 7.888e-05 & $\cdots$ & 3.874e-03 & 8.452e-05 & 9.036e-05 \\
578.95 & 684.21 & 2.661e-04 & 6.248e-05 & 8.601e-05 & 1.634e-04 & 8.489e-05 & 1.171e-04 & $\cdots$ & 2.954e-03 & 8.208e-05 & 7.070e-05 \\
684.21 & 789.47 & 1.544e-04 & 5.042e-05 & 8.060e-05 & 1.143e-04 & 6.767e-05 & 9.872e-05 & $\cdots$ & 2.294e-03 & 6.469e-05 & 6.761e-05 \\
789.47 & 894.74 & 4.640e-05 & 2.696e-05 & 6.357e-05 & 1.861e-04 & 8.099e-05 & 7.323e-05 & $\cdots$ & 1.828e-03 & 6.360e-05 & 6.737e-05 \\
894.74 & 1000.00 & 1.147e-04 & 3.851e-05 & 1.301e-04 & 1.781e-04 & 9.041e-05 & 5.314e-05 & $\cdots$ & 1.637e-03 & 6.314e-05 & 6.449e-05 \\
1000.00 & 1105.26 & 1.709e-04 & 1.266e-04 & 5.535e-05 & 1.049e-04 & 4.654e-05 & 1.380e-04 & $\cdots$ & 1.172e-03 & 5.437e-05 & 5.361e-05 \\
1105.26 & 1210.52 & 2.080e-04 & 4.351e-05 & 5.761e-05 & 1.694e-04 & 7.655e-05 & 6.366e-05 & $\cdots$ & 9.701e-04 & 4.994e-05 & 5.215e-05 \\
1210.52 & 1315.78 & 1.635e-04 & 5.292e-05 & 9.524e-05 & 1.968e-04 & 3.395e-05 & 4.878e-05 & $\cdots$ & 9.479e-04 & 5.161e-05 & 5.127e-05 \\
\hline
\end{tabular}
\end{table*}

\label{lastpage}
\end{document}